\documentclass[longauth, printer]{aa}
\pdfoutput = 1
\usepackage[varg]{txfonts}
\usepackage{mathabx}
\usepackage{graphicx}
\usepackage{upgreek}
\usepackage{makecell}
\usepackage{xcolor}

\usepackage{tikz}
\usetikzlibrary{decorations.pathreplacing,decorations.markings,snakes}

\usepackage{float}

\usepackage{layouts}
\usepackage{printlen}

\usepackage{natbib}
\bibpunct{(}{)}{;}{a}{}{,}

\def\ii{{\mathrm{i}}}
\def\jj{{\mathrm{j}}}
\def\ps{{\mathrm{ps}}}
\def\as{{\mathrm{as}}}
\def\pc{{\mathrm{pc}}}
\def\ma{{\mathrm{max}}}
\def\RA{{\mathrm{R.A.}}}
\def\DEC{{\mathrm{Dec.}}}
\def\fib{{\mathrm{fib}}}
\def\tel{{\mathrm{tel}}}
\def\obj{{\mathrm{obj}}}
\def\obs{{\mathrm{obs}}}
\def\mod{{\mathrm{mod}}}
\def\bin{{\mathrm{bin}}}
\def\mas{{\mathrm{mas}}}
\def\bkg{{\mathrm{bkg}}}
\def\central{{\mathrm{cent}}}

\def\pup{{\mathrm{pup}}}
\def\foc{{\mathrm{foc}}}
\def\turb{{\mathrm{turb}}}
\def\estat{{\mathrm{stat.}}}
\def\esys{{\mathrm{sys.}}}
\def\ft{\mathcal{F}}
\newcommand{\av}[1]{{\left\langle #1 \right\rangle}}

\title{Improved GRAVITY astrometric accuracy from modeling of optical aberrations}
   \author{GRAVITY Collaboration\thanks{GRAVITY is developed
    in a collaboration by the Max Planck Institute for
    extraterrestrial Physics, LESIA of Observatoire de Paris/Universit\'e PSL/CNRS/Sorbonne Universit\'e/Universit\'e de Paris and IPAG of Universit\'e Grenoble Alpes /
    CNRS, the Max Planck Institute for Astronomy, the University of
    Cologne, the CENTRA - Centro de Astrofisica e Gravita\c c\~ao, and
    the European Southern Observatory. $\,\,\,\,\,\,$ Corresponding authors: J.Stadler (email jstadler@mpe.mpg.de) and F. Widmann (email fwidmann@mpe.mpg.de).
    }:
R.~Abuter\inst{8}
\and A.~Amorim\inst{6,12}
\and M.~Baub\"ock\inst{1}
\and J.P.~Berger\inst{5,8}
\and H.~Bonnet\inst{8}
\and W.~Brandner\inst{3}
\and Y.~Cl\'{e}net\inst{2}
\and R.~Davies\inst{1}
\and P.T.~de~Zeeuw\inst{10,1}
\and J.~Dexter\inst{13,1}
\and Y.~Dallilar\inst{1}
\and A.~Drescher\inst{1,16}
\and A.~Eckart\inst{4,17}
\and F.~Eisenhauer\inst{1}
\and N.M.~F\"orster~Schreiber\inst{1} 
\and P.~Garcia\inst{7,12}
\and F.~Gao\inst{1}
\and E.~Gendron\inst{2}
\and R.~Genzel\inst{1,11}
\and S.~Gillessen\inst{1}
\and M.~Habibi\inst{1}
\and X.~Haubois\inst{9}
\and G.~Hei{\ss}el\inst{2}
\and T.~Henning\inst{3}
\and S.~Hippler\inst{3}
\and M.~Horrobin\inst{4}
\and A.~Jim\'enez-Rosales\inst{1}
\and L.~Jochum\inst{9}
\and L.~Jocou\inst{5}
\and A.~Kaufer\inst{9}
\and P.~Kervella\inst{2}
\and S.~Lacour\inst{2}
\and V.~Lapeyr\`ere\inst{2}
\and J.-B.~Le~Bouquin\inst{5}
\and P.~L\'ena\inst{2}
\and D.~Lutz\inst{1}
\and M.~Nowak\inst{15,2}
\and T.~Ott\inst{1}
\and T.~Paumard\inst{2}
\and K.~Perraut\inst{5}
\and G.~Perrin\inst{2}
\and O.~Pfuhl\inst{8,1}
\and S.~Rabien\inst{1}
\and G.~Rodr\'iguez-Coira\inst{2}
\and J.~Shangguan\inst{1}
\and T.~Shimizu\inst{1}
\and S.~Scheithauer\inst{3}
\and J.~Stadler\inst{1}
\and O.~Straub\inst{1}
\and C.~Straubmeier\inst{4}
\and E.~Sturm\inst{1}
\and L.J.~Tacconi\inst{1}
\and F.~Vincent\inst{2}
\and S.~von~Fellenberg\inst{1}
\and I.~Waisberg\inst{14,1}
\and F.~Widmann\inst{1}
\and E.~Wieprecht\inst{1}
\and E.~Wiezorrek\inst{1} 
\and J.~Woillez\inst{8}
\and S.~Yazici\inst{1,4}
\and A.~Young\inst{1}
\and G.~Zins\inst{9}
}

\institute{
Max Planck Institute for extraterrestrial Physics,
Giessenbachstra{\ss}e~1, 85748 Garching, Germany
\and LESIA, Observatoire de Paris, Universit\'e PSL, CNRS, Sorbonne Universit\'e, Universit\'e de Paris, 5 place Jules Janssen, 92195 Meudon, France
\and Max Planck Institute for Astronomy, K\"onigstuhl 17, 
69117 Heidelberg, Germany
\and $1^{\rm st}$ Institute of Physics, University of Cologne,
Z\"ulpicher Stra{\ss}e 77, 50937 Cologne, Germany
\and Univ. Grenoble Alpes, CNRS, IPAG, 38000 Grenoble, France
\and Universidade de Lisboa - Faculdade de Ci\^encias, Campo Grande,
1749-016 Lisboa, Portugal 
\and Faculdade de Engenharia, Universidade do Porto, rua Dr. Roberto
Frias, 4200-465 Porto, Portugal 
\and European Southern Observatory, Karl-Schwarzschild-Stra{\ss}e 2, 85748
Garching, Germany
\and European Southern Observatory, Casilla 19001, Santiago 19, Chile
\and Sterrewacht Leiden, Leiden University, Postbus 9513, 2300 RA
Leiden, The Netherlands
\and Departments of Physics and Astronomy, Le Conte Hall, University
of California, Berkeley, CA 94720, USA
\and CENTRA - Centro de Astrof\'{\i}sica e
Gravita\c c\~ao, IST, Universidade de Lisboa, 1049-001 Lisboa,
Portugal
\and Department of Astrophysical \& Planetary Sciences, JILA, Duane Physics Bldg., 2000 Colorado Ave, University of Colorado, Boulder, CO 80309, USA
\and Department of Particle Physics \& Astrophysics, Weizmann Institute of Science, Rehovot 76100, Israel
\and Institute of Astronomy, Madingley Road, Cambridge CB3 0HA, UK
\and Department of Physics, Technical University Munich, James-Franck-Straße 1,  85748 Garching, Germany
\and Max Planck Institute for Radio Astronomy, Auf dem H\"ugel 69, 53121 Bonn, Germany
}

\abstract
{The GRAVITY instrument on the ESO VLTI pioneers the field of high-precision near-infrared interferometry by providing astrometry at the $10 - 100\,\mu\as$ level. Measurements at such high precision crucially depend on the control of systematic effects. Here, we investigate how aberrations introduced by small optical imperfections along the path from the telescope to the detector affect the astrometry. We develop an analytical model that describes the impact of such aberrations on the measurement of complex visibilities. Our formalism accounts for pupil-plane and focal-plane aberrations, as well as for the interplay between static and turbulent aberrations, and successfully reproduces calibration measurements of a binary star. The Galactic Center observations with GRAVITY in 2017 and 2018, when both Sgr A* and the star S2 were targeted in a single fiber pointing, are affected by these aberrations at a level of less than 0.5 mas. Removal of these effects brings the measurement in harmony with the dual beam observations of 2019 and 2020, which are not affected by these aberrations. This also resolves the small systematic discrepancies between the derived distance $R_0$ to the Galactic Center reported previously.
}

\keywords{instrumentation: high angular resolution,  instrumentation: interferometers,  methods: data analysis,  galaxy: center,  galaxy: fundamental parameters}

\begin{document}
\maketitle

\section{Introduction}
\label{sec: introduction}
The distance to the Galactic Center (GC), $R_0$, can be measured directly from stellar orbits around Sgr~A*, the radio source associated with the GC massive black hole (MBH) (see e.g. \citealt{2010RvMP...82.3121G} and \citealt{2016ARA&A..54..529B} for a recent overview of alternative methods). To this end, the star's proper motion, given in angle per unit time, is compared to its radial velocity, obtained in absolute length per units time from spectroscopic observations. The GC distance then follows directly as a scaling parameter between the two measurements. Most suited to measure $R_0$ is S2, a massive young main sequence B-star on a 16-year orbit with semi-major axis $a \simeq 125\,\mas$ and apparent K-band magnitude $m_k \simeq 14$ \citep{2003ApJ...586L.127G, 2005ApJ...628..246E, 2008ApJ...672L.119M, 2009ApJ...707L.114G, 2017ApJ...837...30G, 2017ApJ...847..120H}. During its pericenter passage in 2018, S2 was closely monitored in astrometry and spectroscopy \citep{2018A&A...615L..15G, 2019Sci...365..664D}. In particular, the GRAVITY instrument \citep{2017A&A...602A..94G} directly measured the distance between S2 and Sgr~A* during the fly-by at high angular resolution of around $30\,\mu$as. The combination of ultra-high astrometric precision from near-infrared interferometry and the spectroscopic precision of $\lesssim 10\,$km/s allowed to determine the GC distance at the unprecedented precision of $<1$\% \citep{2019A&A...625L..10G}. 

Operating in the K-band, GRAVITY combines the light from either the four Unit Telescopes (UTs) or Auxiliary Telescopes (AT) of the ESO Very Large Telescope Interferometer (VLTI). Fringe tracking on a bright reference object allows for minute-long integration times on the fainter science target and for the measurement of differential complex visibilities. The instrument's extremely high angular resolution of $\simeq 3\,\mas$ results in very accurate astrometry with error bars between $10\,\mu$as and $100\,\upmu\as$ \citep{2017A&A...602A..94G}. However, the latest $R_0$ measurement in \cite{2020A&A...636L...5G}
indicates a possible systematic difference with earlier determinations \citep{2018A&A...615L..15G, 2019A&A...625L..10G}. While the shift is small, of $\mathcal{O}\left(1\%\right)$ only, it is nevertheless significant due to the high precision of the measurement.

The difference in the measured GC distance coincides with a change in the observing mode. GRAVITY observes the Galactic Center with two different methods, depending on the separation between Sgr~A* and S2. Close to pericenter passage, i.e. in 2017 and 2018, the sources are detected simultaneously in a single fiber pointing in the so-called single-beam mode. In later epochs, their separation exceeds the fiber's field of view (FOV), and S2 and Sgr~A* are targeted individually. This is referred to as dual-beam mode. 

In single-beam mode, it is not possible to align the two sources with the fiber center. Hence, to further improve the GRAVITY astrometry, we conducted an analysis of how optical aberrations affect the visibility measurement across the full field of view. A similar concept of field-dependent errors already exist in radio interferometry, where it is known as direction dependent effects (DDEs) (see e.g. \cite{2008A&A...487..419B, 2011A&A...527A.107S, 2015MNRAS.449.2668S, 2018A&A...611A..87T} and references there in). The DDEs can arise either at the instrument level from the antenna beam pattern or at the atmospheric level such as from the ionosphere. In particular for the latest generation of interferometers (e.g. VLA, Meerkat, LOFAR) with a wide FOV and a large fractional bandwidth DDEs cannot be neglected. However, to our knowledge there is no equivalent discussion in the context of optical/near-IR interferometry. 

Indeed, our analysis shows that small optical imperfections in the beam combiner induce field-dependent phase errors that reflect in the inferred binary separation. We developed an analytical model to describe this effect, and verified it by application to a dedicated test-case observation. Applied to the GC observations, the model induces a shift in the S2 relative position of order $0.1 - 0.2\, \mas$ in 2018 and $\sim 0.5\,\mathrm{mas}$ in 2017 in both right ascension (R.A.) and declination (Dec.). Despite being small, the change is non-negligible at the high astrometric accuracy achieved by GRAVITY. We can show that the corrected 2017 and 2018 data is in harmony with the dual-beam observations of 2019 and 2020. Further, when retroactively applying the correction to the data sets used in \cite{2018A&A...615L..15G, 2019A&A...625L..10G}, the ensuing GC distance is fully consistent with the latest result \citep{2020A&A...636L...5G}.

We introduce the analytical model in  Sec. \ref{sec: formalism} and compare it to calibration measurements in Sec. \ref{sec: characterization}. Verification from the binary test-case and the improved S2 position are presented in Sec. \ref{sec: application}, while we discuss the implications for the GC distance in Sec. \ref{sec: results}. Finally, we conclude in Sec. \ref{sec: discussion}. 

\section{Formal description of static aberrations and their impact on visibility measurements}
\label{sec: formalism}
Static aberrations along the instrument's optical path affect the measured visibilities by introducing a complex, field-dependent factor for each telescope. We express this gain in its polar representation and decompose it into a phase map $\phi_i \left(\vec{\alpha}\right)$ and an amplitude map $A_i\left(\vec{\alpha}\right)$. Here, the index $i$ labels the telescope and $\vec{\alpha}$ denotes positions in the image plane. Phase and amplitude maps lead to a modification of the observed complex visibilities $V^\mathrm{obs}$ from the well-known van Cittert-Zernike theorem (c.f. Eq. \ref{eq: visibilities-theory-general}). As we demonstrate in the following, they are given by
\begin{equation}
V^\mathrm{obs} = \frac{\int d\vec{\alpha}\, A_{i}\left(\vec{\alpha}\right)A_{j}\left(\vec{\alpha}\right) O\left(\vec{\alpha}\right) e^{-2\pi i\, \vec{\alpha}\cdot \vec{b}_{i,j}/\lambda + i\left(\phi_i\left(\vec{\alpha}\right) - \phi_j\left(\vec{\alpha}\right)\right)} }
{\sqrt{\int d\vec{\alpha}\, A_i^2 \left(\vec{\alpha}\right) O\left(\vec{\alpha}\right) ~\int d\vec{\alpha}\, A_j^2 \left(\vec{\alpha}\right) O\left(\vec{\alpha}\right)}}\,,
\label{eq: overlap-vCZ-generalized}
\end{equation}
where $\vec{b}_{i,j}$ is the baseline vector between the two telescopes and $O\left(\vec{\alpha}\right)$ denotes the intensity distribution of the observed astronomical object.

In this section, we show how the phase- and amplitude-maps follow from optical aberrations. To this end, we start from the overlap integral, which determines the electromagnetic field from a single telescope arriving at the beam combiner. Subsequently, we propagate the effect of static aberrations from the overlap integral to the measured complex visibility to arrive at a rigorous derivation of Eq.~(\ref{eq: overlap-vCZ-generalized}). Finally, we account for the superposition of static and turbulent aberrations, to obtain a formalism which is applicable in realistic observation scenarios.
\subsection{Static, field-dependent aberrations at fiber injection}
\label{sec: aberrations}
Single mode fibers transport the light collected by each telescope $E_\tel$ to the beam combiner instrument. The overlap integral between light and the fiber mode $E_\fib$ then determines the transmitted electric field \citep{1988smf..book.....N},
\begin{equation}
E\left(\vec{\beta}\right) = E_\fib\left(\vec{\beta}\right)\times \eta
=  E_\fib \times  \int d\vec{\xi}~ E_\tel\left(\vec{\xi}\right) E^*_\fib\left(\vec{\xi}\right)\,.
\end{equation}
Here, we assume a normalized fiber mode $\int d\vec{\xi} \left|E_\fib\left(\vec{\xi}\right) \right|^2 = 1$ and express image-plane positions by  two-dimensional vectors, $\vec{\xi}$ and $\vec{\beta}$. Following the description of \cite{2019A&A...625A..48P}, the overlap integral is converted to the pupil plane by the Parseval-Plancharel theorem,
\begin{equation}
\eta =\int d\vec{u}~ \ft^{-1}\left[E_\obj\right] P\left(\vec{u}\right)\, \ft^{-1}\left[{E}_\fib^*\right]\left(\vec{u}\right)\,,
\label{eq: overap-pupil-plane-intergral}
\end{equation}
where $\mathcal{F}^{-1}$ denotes the inverse Fourier transform, i.e. transformation from the image to the pupil plane, and $E_\obj$ the light emitted by the astronomical object. The latter is connected to $\ft^{-1}\left(E_\tel\right)$ by multiplication with the pupil function $P\left(\vec{u}\right)$, corresponding to a convolution in the image plane. In the most simple case of a single point source located at $\vec{\alpha}_0$, the light is described by a pure phase $\ft^{-1}[E^\ps_\obj]=\exp\left(-2\pi i\, \vec{u}\cdot\vec{\alpha}_0\right)$. The pupil- and image-plane coordinates, $\vec{\xi}$ and $\vec{u}$ respectively, are Fourier-conjugate to each other and chosen to be dimensionless. That is, any length scale in the pupil plane is given by $\lambda u$ where $\lambda$ refers to the wavelength and $u = \left|\vec{u}\right|$. For discussion, we convert the dimensionless image plane coordinates $\vec{\xi}$ to the corresponding angular separation in UT observations. In an aberration-free scenario, the pupil function of a spherical telescope with diameter $2r_\tel$ and central obscuration $2r_\central$ simply is
\begin{equation}
\tilde{P}\left(\vec{u}\right) =
\begin{cases}
0 &\quad\mathrm{if}\quad u \leq r_\central/\lambda\\
1 &\quad\mathrm{if}\quad r_\central < u \leq r_\tel/\lambda\\
0 &\quad\mathrm{if}\quad u > r_\tel/\lambda 
\end{cases}\,.
\label{eq: overlap-ppupil}
\end{equation}
Optical aberrations multiply the pupil function by a position-dependent, complex phase, and we here consider the case of purely static aberrations. These are characterized by an optical path difference (OPD) $d_\pup\left(\vec{u}\right)$ in the pupil plane that can be expanded in terms of Zernike polynomials $Z_n^m$,
\begin{equation}
d_\pup \left(\vec{u}\right) = \sum_{n=0}^{n_\ma} \sum_{m=-n}^{n} A_n^m\, Z_n^m\left(\lambda\vec{u}/r_\tel\right)\,.
\label{eq: zernike-decomposition}
\end{equation}
We adopt the convention that $Z_n^m$ is dimensionless and the coefficient $A_n^m$ corresponds to the term's root mean square over the unit circle. Defining the turbulence-free complex fiber mode apodised by the pupil function as 
\begin{equation}
\Pi_\circledcirc = e^{2\pi i \, d_\pup\left(\vec{u}\right)/\lambda}\, \tilde{P}\left(\vec{u}\right)\, \ft^{-1}\left[{E}_\fib^*\right]\left(\vec{u}\right)\,,
\label{eq: overlap-complex-pupil-function}
\end{equation}
the overlap integral reads

\begin{equation}
\eta = \int d\vec{u}~ \ft^{-1}\left[E_\obj\right]\left(\vec{u}\right)\, \Pi_\circledcirc\left(\vec{u}\right)\,.
\label{eq: overlap-final}
\end{equation}

The overlap integral obviously depends on the fiber profile which, for a perfectly aligned ideal single-mode fiber, is
\begin{equation}
\ft^{-1}\left[\tilde{E}_\fib^*\right] = \exp\left(-\frac{\lambda^2 u^2}{2\,\sigma_\fib^2}\right)\,.
\label{eq: overlap-ideal-fiber-profile}
\end{equation}
GRAVITY was designed for optimal fiber injection \citep{2014SPIE.9146E..23P}, which is obtained for $\sigma_\fib = 2r_\tel\sqrt{2\ln 2}/\left(\pi\epsilon\right)$ \citep{2002JOSAA..19.2445W}. Here, the parameter $\epsilon$ is of order unity and describes the pupil shape. 

From comparison between model predictions and the calibration measurements in Sec.~\ref{sec: zernike}, we find that pupil-plane distortions alone are not sufficient to describe the observed aberration pattern. We also need to account for optical errors in the focal plane. Misalignment of the optical fiber, as well as higher order aberrations at fiber injection, introduce a complex phase to Eq.~(\ref{eq: overlap-ideal-fiber-profile}) and can distort the amplitude of the fiber profile. 

\begin{figure}
\begin{tikzpicture}
\draw [fill=gray, draw=gray] (2.,0.) rectangle (2.2, -.5);
\draw [fill=gray, draw=gray] (2.,-3.5) rectangle (2.2, -4);
\draw[snake, thick] (2.1, -0.8) -- (2.1, -3.3);
\node[anchor=north west, inner sep=0pt] at (2.5, -.5) (np) {$d_\pup\left(\vec{u}\right)$};
\draw[->] (np.south) to[out=-90, in=0] ++(-.8,-.5);
\begin{scope}[decoration={
    markings,
    mark=at position 0.5 with {\arrow{>}}}
    ] 
    \draw[postaction={decorate}] (1.0, -1.) -- (2.1, -1.);
    \draw[postaction={decorate}] (1.0, -3.) -- (2.1, -3.);
    \draw[postaction={decorate}, ->] (2.1, -1) -- (6.5, -2.5);
    \draw[postaction={decorate}, ->] (2.1, -3) -- (6.5, -1.5);
\end{scope}

\node[anchor=west, text width=2.5cm, minimum height=1.8cm, fill=gray!50, rotate=20, inner sep=0pt, draw=gray!50] (n1) at (7., -2.4) {};
\draw[gray!50, inner sep=0pt] (n1.north west) -- (n1.south west) node[inner sep=1pt, fill=black, circle, pos=0.5] (n2) {};
\path[fill=gray!50, draw=gray!50, inner sep=0pt, line width=0.0pt] (n1.north east) -- (n1.south east|- -20:1.4) -- (n1.south east) -- (n1.north east);
\draw[snake=coil,segment aspect=0, thick] (n1.north west) -- ++(-70: .6cm);
\draw[snake=coil,segment aspect=0, thick] (n1.south west) -- ++(110: .88cm);
\node[anchor=west, xshift=3.cm] at (np.east) (nf) {$d_\foc\left(\vec{x}\right)$};
\draw[->] (nf.south) to[out=-90, in=0] ++(-.4,-.7);
\node[anchor=south, inner sep=0pt, rotate=20] at ([xshift=-.2cm, yshift=-.08cm]n1.north east) {optical fiber};

\draw[<->] (n2) -- ++(0cm, .4cm) node[midway, anchor=west, yshift=-.1cm] (n3) {$\delta x = f\cdot \xi_\fib$};
\node[anchor=west, circle, draw=black, inner sep=1pt, xshift=-.2cm, yshift=.2cm] at (n3.east) {A};
\draw[very thin, dashed](n1.south west) -- ++(2.5cm,0cm);
\draw[] (2.1, -3.7) -- (5., -3.7) node[midway, fill=white] {f};
\draw[] (5., -3.7) -- (6.8, -3.7)node[midway, fill=white] (n5) {$\delta z$};
\node[anchor=south, inner sep=1pt, xshift=-.4cm, yshift=-.2cm, circle, draw=black] at (n5.north) {C};
\draw(2.1, -3.75) -- (2.1, -3.65);
\draw(5., -3.75) -- (5., -3.65);
\draw(6.8, -3.75) -- (6.8, -3.65);
\draw ([xshift=1.8cm] n1.south west) to[bend right] node[midway, xshift=-.4cm, yshift=-.1cm] (n4) {$\varphi$} ++(0cm, .7cm);
\node[anchor=north, inner sep=1pt, xshift=-.4cm, yshift=.2cm, circle, draw=black, fill=white] at (n4.south) {B};
\draw[dashed] (1.0, -2) -- (7.8, -2);

\end{tikzpicture}
\caption{Schematic depiction of the pupil and focal plane aberrations which enter the overlap integral. Both effects in combination are required to describe the aberration patterns observed in calibration measurements. The lowest-order aberrations in the pupil function are shown explicitly, which are (A) lateral fiber misplacement, (B) fiber tilt and (C) defocus. Their effect is further explained in the text. }
\label{fig: overlap-fiber-misalignment}
\end{figure}
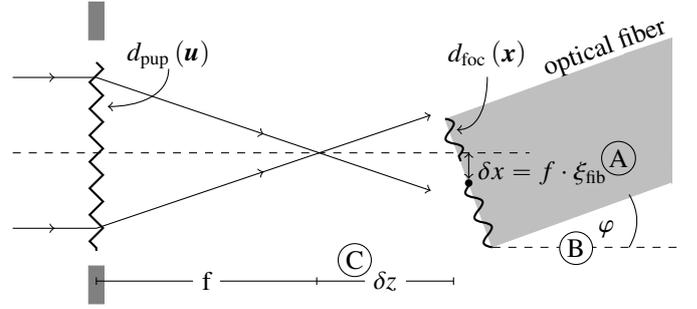

To illustrate the effect of focal plane aberrations, we first consider the three types of misalignment depicted in Fig.~\ref{fig: overlap-fiber-misalignment}: (A) Lateral misplacement of the fiber by $\left(\delta x, \delta y\right)$, which in the pupil plane produces a phase slope $\vec{\xi}_\fib = \left(\delta x/f,\, \delta y/f\right)$, with $f$ being the focal length. (B) Fiber tilt by an angle $\vec{\varphi}_\fib = (\varphi_1, \varphi_2)$ with respect to the optical axis of the system  which shifts the back-propagated fiber mode by $\vec{u}_\fib = \vec{\varphi}\cdot f/\lambda$. And (C), a defocus or axial fiber misplacement by $\delta z$ that introduces an additional phase curvature $\exp\left[\pi\,i\delta z \lambda/f^2\, u^2\right]$. Taking all three effects into account, the generalized fiber profile, projected to the pupil, is \citep{2002JOSAA..19.2445W}
\begin{align}
\ft^{-1}\left[{E}_\fib^*\right] &= \ft^{-1}\left[\tilde{E}_\fib^*\right]\left(\vec{u} - \vec{u}_\fib\right) \\
&\times \exp\left\{-2\pi\,i\left[\frac{\pi\delta z}{2 f^2} \left(\vec{u} -\vec{u}_\fib\right)^2 -   \vec{\xi}_\fib\cdot\left(\vec{u}-\vec{u}_\fib\right)\right]\right\}\,. \nonumber
\end{align}
By rearranging the phase term in the pupil plane, one can decompose it into a piston, tip-tilt and defocus
\begin{align}
d_\fib^\mathrm{piston}\left(\vec{u}\right) &= -\lambda\left(\frac{\delta z \lambda}{f^2}\, \vec{u}_\fib + \vec{\xi}_\fib\right) \cdot \vec{u}_\fib - \frac{\delta z}{4 f^2}\,,
\label{overlap-focal-opd1}\\
\label{eq: overlap-fiber-piston}
d_\fib^\mathrm{tip-tilt} \left(\vec{u}\right) &= \lambda\left(\frac{\delta z \lambda}{f^2} \vec{u}_\fib + \vec{\xi}_\fib\right) \cdot \vec{u}\,,\\
d_\fib^\mathrm{defocus}\left(\vec{u}\right) &= -\frac{\delta z }{4 f^2} \left(2\lambda^2 \vec{u}^2 - 1\right)\,.
\label{eq: overlap-fiber-defocuss}
\end{align}
The phase terms in Eqs. (\ref{overlap-focal-opd1}) to (\ref{eq: overlap-fiber-defocuss}) thus affects the overlap integral in the same way as the lowest-order aberrations in $d_\pup\left(\vec{u}\right)$. For the coordinate shift of the Gaussian profile, on the other hand, there is no such correspondence, and it alters the way in which the optical fiber scans the pupil-plane aberrations.

During GRAVITY observations, the misplacement term (A) depends on the performance of the fiber tracker but also on the uncertainty of the source position. In particular for exoplanet observations, the latter can be sizable. Fiber tilt (B) is controlled by the GRAVITY pupil tracker, and the adaptive optics calibration is one example that impacts the defocus (C).

While lateral misplacement (A) and defocus (C) describe the misplacement of a point-like fiber entrance, fiber tilt (B) accounts for the alignment of the fiber's surface. This surface can exhibit irregularities beyond a simple tilt, which lead to a position-dependent OPD in the focal plane, $d_\foc\left(\vec{x}\right)$, as illustrated in Fig.~\ref{fig: overlap-fiber-misalignment}. Generally, aberrations from optical elements not conjugated to the pupil are field-dependent and known as Seidel aberrations. In this context, $d_\foc\left(\vec{x}\right)$ arising in the focal plane constitutes an extreme example. Still, it is possible to decompose the focal plane distortions into a series of Zernike polynomials, in analogy to Eq.~(\ref{eq: zernike-decomposition}). In this representation, axial fiber offset (C) and fiber tilt (B) simply correspond to the lowest-order coefficients, and higher-order terms amount to a generalization of \cite{2002JOSAA..19.2445W}. Again, the phase terms introduced in $\ft^{-1}\left[\tilde{E}_\fib^*\right]$ by higher order aberrations are degenerate with $d_\pup\left(\vec{u}\right)$, but the amplitude distortions need to be modeled explicitly by themselves.
\begin{figure*}
\begin{center}
\includegraphics[]{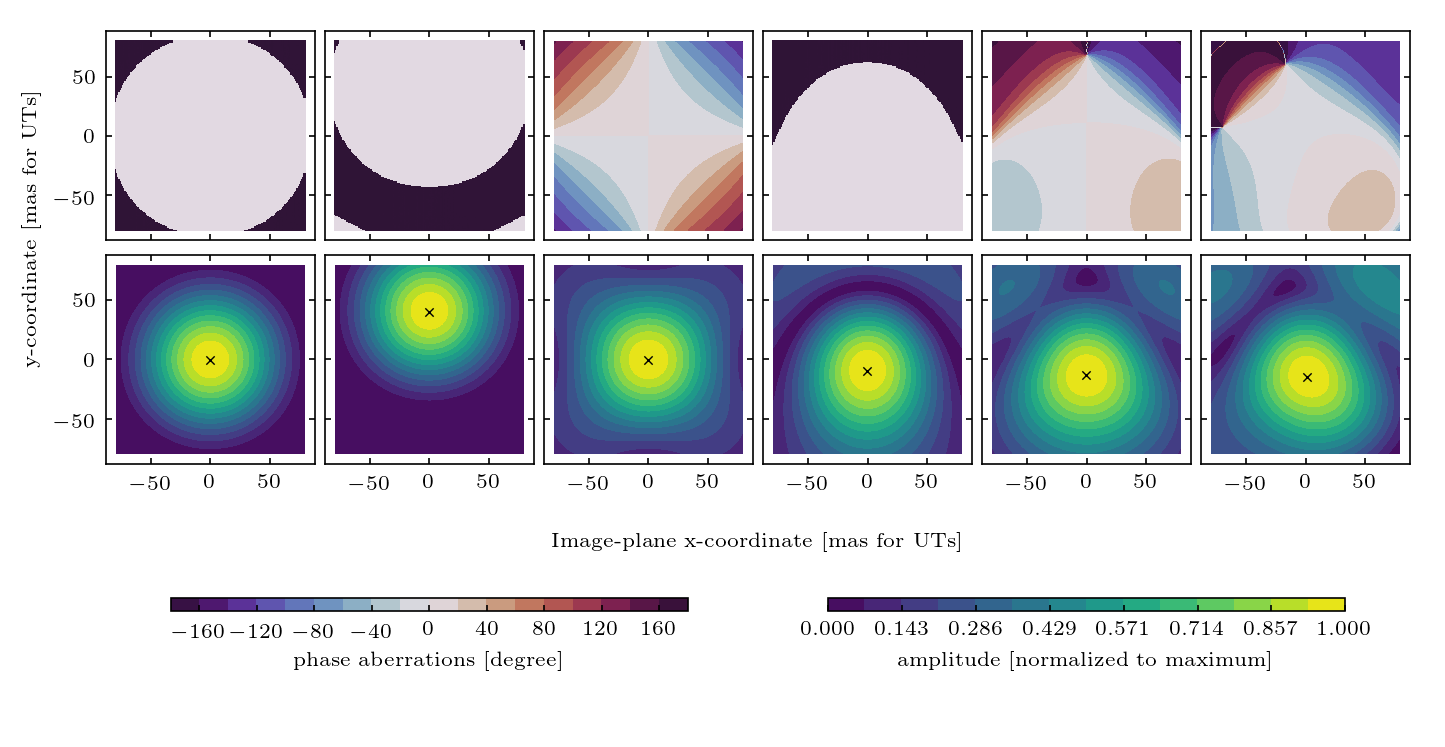}
\caption{Example phase screens (top) and amplitude maps (bottom) in the image plane induced by low-order Zernike aberrations in the pupil plane at a wavelength of $\lambda_0 = 2.2\,\mu\mathrm{m}$. From left to right the considered aberrations are: perfect Airy pattern, vertical tilt of $0.4\, \upmu\mathrm{m}$ RMS, vertical astigmatism of $0.2\, \upmu\mathrm{m}$ RMS, vertical coma of $0.2\, \upmu\mathrm{m}$ RMS, and combination of astigmatism, coma and trefoil (with RMS $0.2\, \upmu\mathrm{m}$, $0.2\, \upmu\mathrm{m}$, and $0.1\, \upmu\mathrm{m}$, respectively). The rightmost panel also considers an additional fiber tilt with $0.2\, \upmu\mathrm{m}$ RMS.}
\label{fig: example-maps}
\end{center}
\end{figure*}

Finally, for a single point source, located at $\vec{\alpha}_0$ in the image plane, the overlap integral averaged over a time scale much longer than the source's coherence time $\av{...}_\obj$ is
\begin{equation}
\av{\eta^\ps}_\obj \propto \int d\vec{u}~ e^{-2\pi i \,\vec{u}\cdot\vec{\alpha}_0} \Pi_\circledcirc\left(\vec{u}\right)
= \ft\left[\Pi_\circledcirc\right]\left(\vec{\alpha}_0\right)\,.
\label{eq: overlap-pointsource}
\end{equation}
Evaluation of the Fourier transform as function of $\vec{\alpha}_0$ results in a two-dimensional complex map. We show several examples of such maps in Fig.~\ref{fig: example-maps}, assuming different Zernike coefficients to determine $d_\pup\left(\vec{u}\right)$. The perfect Airy pattern, obtained in the limit of zero aberrations, exhibits zero phase in the central part and a phase jump by $180^\circ$ at $\left|\vec{\alpha}\right| \simeq 1.22\,\lambda/\left(2r_\tel\right)$. Anti-symmetric terms, such as tilt, coma and trefoil (not shown), only alter the location and shape of the phase jump, while defocus (not shown), astigmatism and higher order terms produce smooth phase gradients. For a general choice of $d_\pup\left(\vec{u}\right)$ and in the absence of focal-plane aberrations, there is a saddle point where the phase maps average to zero, but significant phase shifts are encountered at larger radii. 

Focal-plane aberrations break the radial symmetry of the fiber profile. Still, if the perturbations are small enough, the phase maps show a saddle point, but its value differs from zero and its location may be shifted. In any case, the transmitted amplitude is deformed and/or misplaced from the perfect Airy case. Pupil-plane aberrations typically widen the amplitude, while image-plane aberrations have the opposite effect. They lead to a widening of the fiber in the pupil plane and correspondingly to a narrower image-plane profile. The exact scaling relation for the position of the Airy ring remains true only approximately in the presence of higher-order aberrations such that maps at two different wavelengths, $\lambda_1$ and $\lambda_2$, can be related by
\begin{equation}
\av{\eta^\ps}_\obj\left(\alpha_0, \lambda_1\right) \simeq \av{\eta^\ps}_\obj\left(\alpha_0\, \frac{\lambda_2}{\lambda_1},\, \lambda_2\right)\,.
\label{eq: overlap-approximate-scaling}
\end{equation}

\subsection{Effect on visibility measurements and astrometry}
\label{sec: astrometry}
The overlap integral defines the electromagnetic wave transmitted to the beam combiner from each of the four telescopes. After pairwise beam combination, the complex visibilities are obtained from the inference pattern $I_{i,j}$,
\begin{align}
I_{i,j} &= \int d\vec{\beta}~ \av{\left|E_i\left(\vec{\beta}\right) + E_j\left(\vec{\beta}\right)\right|^2}_\obj\\
&= \av{\left|\eta_i\right|^2}_\obj + \av{\left|\eta_j\right|^2}_\obj + 2\Re\, \av{\eta_i\eta_j^*}_\obj\,,
\end{align}
where $i$ and $j$ denote the telescopes involved in the measurement and $I$ is the intensity. The complex pupil function enters each of these terms. Focusing on the single-telescope component first, we find from Eq.~(\ref{eq: overlap-final})
\begin{align}
\av{\left|\eta_i\right|^2}_\obj &= \int d\vec{\alpha}~ \ft\left[ \Pi_{\circledcirc,i} \otimes \Pi_{\circledcirc,i}\right]\left(\vec{\alpha}\right)\, O\left(\vec{\alpha}\right) \nonumber \\
&= \int d\vec{\alpha}~ \left\vert\ft\left[\Pi_{\circledcirc,i}\right]\left(\vec{\alpha}\right)\right\vert^2  O\left(\vec{\alpha}\right)\,,
\end{align}
where the $\otimes$-operator denotes auto-correlation, and $O\left(\vec{\alpha}\right) = \left|E_\mathrm{obj}\left(\vec{\alpha}\right)\right|^2$ is the brightness distribution of the observed astronomical object which obeys
\begin{equation}
\av{\ft^{-1}\left[E_\obj\right]\left(\vec{u}\right)\,\ft^{-1}\left[E_\obj\right]^*\left(\vec{v}\right)}_\obj
= \ft^{-1}\left[O\left(\vec{\alpha}\right)\right]\left(\vec{u}-\vec{v}\right)\,.
\end{equation} 
Similarly, the inference term is given by
\begin{align}
&\av{\eta_i\eta_j^*}_\obj = \int d\vec{\alpha}~ \ft\left[\Pi_{\circledcirc,i} \otimes \Pi_{\circledcirc,j}\right]\left(\vec{\alpha}\right)\, O\left(\vec{\alpha}\right) e^{-2\pi i\, \alpha\cdot \vec{b}_{i,j}/\lambda} \nonumber\\
&= \int d\vec{\alpha}~ \ft\left[\Pi_{\circledcirc,i}\right]\left(\vec{\alpha}\right)\, \ft\left[\Pi_{\circledcirc,j}\right]^*\left(\vec{\alpha}\right) O\left(\vec{\alpha}\right) e^{-2\pi i\, \alpha\cdot \vec{b}_{i,j}/\lambda}\,,
\end{align}
where $\vec{b}_{i,j}$ is the baseline vector. 

All optical aberrations discussed previously are encoded in the back-projected apodized pupil, which is a complex field-dependent function. Expressing the pupil function in its polar representation, 
\begin{equation}
\ft\left[ \Pi_{\circledcirc,i}\right] = A_i\left(\vec{\alpha}\right) e^{i\phi_i\left(\vec{\alpha}\right)}\,,
\end{equation}
we refer to $A_i$ as the telescope-dependent "amplitude map" and to $\phi_i$ as the "phase map". Note that these quantities are closely related to the photometric and the interferometric lobes, $L_i\left(\vec{\alpha}\right) = A_i^2\left(\vec{\alpha}\right)$ and 
\begin{align}
L_\mathrm{i,j}\left(\vec{\alpha}\right) 
= A_\mathrm{i}\left(\vec{\alpha}\right) e^{i\phi_\mathrm{i}\left(\vec{\alpha}\right)}\, A_\mathrm{j}\left(\vec{\alpha}\right) e^{-i\phi_\mathrm{j}\left(\vec{\alpha}\right)}\,,
\label{eq: visibilities-phase-amplitude-map-definition}
\end{align}
respectively.


From the measured inference pattern, the complex visibilities are obtained as
\begin{equation}
V^\obs\left(\vec{b}_{i,j}/\lambda\right) 
= \av{\eta_i\eta_j^*}_\obj\bigg/ \sqrt{\av{\left|\eta_i\right|^2}_\obj \av{\left|\eta_j\right|^2}_\obj}\,.
\label{eq: visibilities-measurement-general}
\end{equation}
By contrast, in an ideal, aberration-free setting, the van-Cittert-Zernike theorem relates the complex visibilities to the object's brightness distribution
\begin{equation}
V^\mod\left(\vec{b}_{i,j}/\lambda\right) = \frac{
\int d\vec{\alpha}~ O\left(\vec{\alpha}\right) e^{-2\pi i\, \alpha\cdot \vec{b}_{i,j}/\lambda}
}{
\int d\vec{\alpha}~ O\left(\vec{\alpha}\right)
}\,.
\label{eq: visibilities-theory-general}
\end{equation}
Comparison of Eq. (\ref{eq: visibilities-measurement-general}) and Eq. (\ref{eq: visibilities-theory-general}) readily suggests that static aberrations at fiber injection distort both the measured visibility phases and amplitudes. We thus need to adapt the interferometric equation accordingly. To make this effect even more explicit, we first consider the case of a single, unresolved object at position $\vec{\alpha}_0$,
\begin{equation}
V^{\obs}_\ps\left(\vec{b}_{i,j}/\lambda\right)
=\frac{L_{i,j}\left(\vec{\alpha}_0\right)}{\sqrt{L_i\left(\vec{\alpha}_0\right) L_j\left(\vec{\alpha}_0\right)}}\, e^{-2\pi i \vec{\alpha}_0\cdot\vec{b}_{i,j}/\lambda}\,.
\label{eq: visibility-observed-pointsource}
\end{equation}
In the aberration-free case, the phase and amplitude maps of either telescope are given by the perfect Airy pattern shown in the very left panel of Fig. \ref{fig: example-maps}, and $\phi_{i/j}\left(\vec{\alpha}_0\right)$ equals zero or $2\pi$. The presence of static aberrations introduces a phase shift by $\phi_i\left(\vec{\alpha}_0\right) - \phi_j\left(\vec{\alpha}_0\right)$. For an interferometric binary with positions $\vec{\alpha}_1$, $\vec{\alpha}_2$ and flux ratio $f^\bin$ the measured visibility becomes
\begin{equation}
V^{\obs}_\bin 
=\frac{
L_{i,j}\left(\vec{\alpha}_1\right) e^{-2\pi i \vec{\alpha}_1\cdot\vec{b}_{i,j}/\lambda}
+
f^\bin L_{i,j}\left(\vec{\alpha}_2\right) e^{-2\pi i \vec{\alpha}_2\cdot\vec{b}_{i,j}/\lambda}
}
{\sqrt{
\left[L_i\left(\vec{\alpha}_1\right) + f^\bin L_i\left(\vec{\alpha}_2\right)\right]
\left[L_j\left(\vec{\alpha}_1\right) + f^\bin L_j\left(\vec{\alpha}_2\right)\right]
}} \,.
\label{eq: visibility-observed-binary}
\end{equation}
Finally, for a generic extended object with an intensity distribution $O\left(\vec{\alpha}\right)$ the van-Cittert-
Zernike theorem generalizes to the expression stated at the beginning of this section, in Eq.~(\ref{eq: overlap-vCZ-generalized})
\begin{equation*}
V^\mathrm{obs} = \frac{\int d\vec{\alpha}\, L_{i,j}\left(\vec{\alpha}\right) O\left(\vec{\alpha}\right) e^{-2\pi i\, vec{\alpha}\cdot \vec{b}_{i,j}/\lambda}}{\sqrt{\int d\vec{\alpha}\, L_i \left(\vec{\alpha}\right) O\left(\vec{\alpha}\right) ~\int d\vec{\alpha}\, L_j \left(\vec{\alpha}\right) O\left(\vec{\alpha}\right)}}\,.
\end{equation*}

Single point sources typically are observed at the fiber center, where fiber injection is highest and the phase distortions are close to zero. In situations where a very precise alignment is not possible, like for example in exoplanet observations, the visibilities can pick up some small contribution from the phase maps. For binaries with a separation comparable to the fiber width, a configuration in which the phase and amplitude maps are irrelevant cannot be obtained in principle. In this case, the effect of static aberrations needs to be modeled and corrected for in the data analysis.

\subsection{Interplay with turbulent aberrations}
\label{sec: turbulences}
To this point, we have not considered the effect of time varying phase aberrations. These are introduced by atmospheric turbulence or time-varying imperfections in the optical system such as tip-tilt jitter from the adaptive optics. Their effect is to multiply the static pupil function by another, time dependent phase
\begin{equation}
P_\circledcirc = \Pi_\circledcirc \, e^{i\phi^\turb\left(\vec{u}, t\right)}\,.
\end{equation}
To see how time-dependent aberrations affect the visibility measurement, we briefly recap the arguments of \cite{2019A&A...625A..48P}. Assuming that the detector integration time by far exceeds the coherence time of phase fluctuations, the long-time average $\av{...}_\turb$ over the telescope lobes is
\begin{align}
\av{L_\mathrm{i}\left(\vec{\alpha}\right)}_\turb
& = \av{\left|\ft\left[P_{\circledcirc,i}\right]\left(\vec{\alpha}\right)\right|^2} 
= \nonumber\\ &= 
\ft \left[\left(\Pi_{\circledcirc,\mathrm{i}} \otimes \Pi_{\circledcirc,\mathrm{i}}\right) \left(\vec{u} \right)
e^{-\frac{1}{2}D_\phi\left(\vec{u}\right)} \right]\,,
\\
\av{L_\mathrm{i,j}\left(\vec{\alpha}\right)}_\turb 
&= 
\av{\ft\left[P_{\circledcirc,\mathrm{i}}\right]\left(\vec{\alpha}\right)}_\turb
\av{\ft\left[P_{\circledcirc,\mathrm{j}}\right]\left(\vec{\alpha}\right)}_\turb^*
\nonumber \\ &=
\ft \left[\left(\Pi_{\circledcirc,\mathrm{i}} \otimes \Pi_{\circledcirc,\mathrm{j}}\right) \left(\vec{u} \right)
e^{-\sigma_\phi} \right]\,,
\end{align}
where $D_\phi\left(\vec{u}\right)$ is the structure function of the turbulent phase \citep{1981PrOpt..19..281R}, which saturates to $2\sigma_\phi$ on large scales. Two assumptions underlie these expressions, first that the fluctuations are stationary and second that the baseline between the telescopes is long enough for the respective apertures to become uncorrelated. As in \cite{2019A&A...625A..48P}, we assume both to be fulfilled.

In the case of GRAVITY observations, atmospheric phase variations across the telescope apertures are corrected by the adaptive optics system and the turbulent aberrations are dominated by tip-tilt jitter. Thus, the turbulent phase is
\begin{equation}
\phi^\turb_\mathrm{i} = 2\pi\, \vec{t}_\ii(t) \cdot \vec{u}\,,
\end{equation}
where the two directions of $\vec{t}_\ii(t)$ are independent and follow a Gaussian distribution with zero mean and variance $\sigma_t^2$. The structure function then becomes $D_t\left(\vec{u}\right) = \left(2\pi \sigma_t u\right)^2$, and the photometric lobe is given by
\begin{equation}
\av{L_\mathrm{i}\left(\vec{\alpha}\right)}_\turb =
\left|\ft\left[\Pi_{\circledcirc,\ii}\right]\left(\vec{\alpha}\right)\right|^2 \circledast \exp\left(- \frac{\alpha^2}{2\,\sigma_t^2}\right)\,,
\label{eq: smoothing-photometric-lobe}
\end{equation}
where $\circledast$ denotes convolution. In case of the interferometric lobe, we further assume that the jitter is uncorrelated between telescopes which yields
\begin{equation}
\av{L_\mathrm{i,j}\left(\vec{\alpha}\right)}_\turb  = \left(\ft\left[\Pi_{\circledcirc,i}\right] \circledast e^{-\frac{\alpha^2}{2\sigma^2_t}}\right)^*\left(\ft\left[\Pi_{\circledcirc,j}\right] \circledast e^{-\frac{\alpha^2}{2\sigma^2_t}}\right)\,.
\label{eq: smoothing-interferometric-lobe}
\end{equation}
These turbulent lobes replace the static expressions of the previous sections in the prediction of the observed visibility, i.e. in Eq. (\ref{eq: overlap-vCZ-generalized}), Eq. (\ref{eq: visibility-observed-pointsource}) and Eq. (\ref{eq: visibility-observed-binary}). The tip-tilt jitter acts like a Gaussian convolution kernel on the static maps, which is applied to the amplitude map squared in case of the photometric lobe but to the full complex map in the case of the interferometric lobe.

\section{Measurement and characterization of aberrations for the GRAVITY beam combiner}
\label{sec: characterization}
GRAVITY observes the Galactic Center in its so-called dual-field mode, which requires the presence of a bright reference target (IRS 16C) within 2'' of the actual science targets, Sgr A* and S2. The field at each telescope is split, and reference and science source are separately injected into the fringe tracking (FT) and science channel (SC) fibers. Short detector integration times on the FT allow for the optical path delay to be constantly adjusted for atmospheric turbulence in order to maintain a high fringe contrast. The science channel then measures a differential visibility phase with respect to the fringe tracker on each baseline. 

Phase and amplitude maps are inherently single-field effects in the sense that they individually affect the fringe tracker and the science channel for each telescope separately. Based on the optical layout of the fiber coupler \citep{2014SPIE.9146E..23P}, there is no reason to expect equal aberrations on the SC and FT. However, the fringe tracking object is a bright, unresolved source which is actively tracked by the fiber center in closed loop, such that the phase distortions introduced from static aberrations are small. Moreover, any possible phase distortion from the fringe tracker cancels in the analysis of closure phases or induces a global shift without affecting the binary separation in the analysis of visibility phases. However, a description of the SC phase and amplitude maps is essential to robustly measure a binary separation in the science channel.

Here we report on measurements with the GRAVITY Calibration Unit \citep{2014SPIE.9146E..1UB} and on our subsequent analysis to extract SC phase and amplitude maps. We then fit the static-aberration model from Sec. \ref{sec: aberrations} to those maps in order to demonstrate its validity and to obtain compressed representation of the aberrations in form of a small number of Zernike coefficients.

\subsection{Phase map measurements with the calibration unit}
\label{sec: measurements}
\begin{figure}
\includegraphics[]{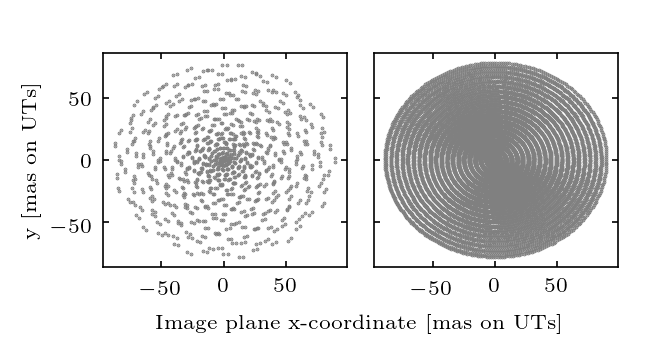}
\caption{Examples of the scanning pattern applied in the Calibration Unit measurements. SC aberration maps where obtained with a slow modulation frequency (left). For the corresponding FT measurement, a faster scanning was used, and the right panel only shows a single iteration of in- and out-spiral.}
\label{fig: measurement-scanning-pattern}
\end{figure}

The GRAVITY Calibration Unit, which we use for the measurement of static aberrations, is directly attached to the beam combiner and creates the light of an artificial science and fringe tracker star. By modulating the voltage on GRAVITY's positioning mirror, the position of that star relative to the fiber can be changed. We scan the FOV out to $\sim 70\,\mathrm{mas}$ in a pattern of in- and out-spiral, which is applied simultaneously to the FT and SC on one single telescope at a time, see Fig.~\ref{fig: measurement-scanning-pattern}.

In normal observation mode, GRAVITY controls the differential OPD between science channel and fringe tracker by its laser metrology and the common path to the telescopes by fringe tracking. During the phase map calibration measurement, however, fringe tracking is not possible because the fringes are lost at the margins of the scanning region. Instead, the common path from the telescope to the instrument drifts in time. Thus the determination of the aberration pattern from the absolute FT and SC phase requires a drift correction. On the FT, the short detector integration time with maximum sampling frequency of $1\,\mathrm{kHz}$ allows one to resolve fast modulation of the source position and the full FOV can be scanned within $\sim 15\,\mathrm{s}$. Over this short time span, the drift is well described by a constant velocity, which we fit and subtract from the data. On the SC, in contrast, the minimum detector integration time is $0.13\,\mathrm{s}$ and a full scan of the FOV takes $2-3\,\mathrm{minutes}$, too long to model the drift by a simple polynomial fit. Instead, we obtain the science channel aberrations via a detour and first analyze the differential, drift-free SC-FT phase. The pure science channel aberrations then follow from knowledge of the absolute fringe tracker phase.

The data are reduced by the standard GRAVITY pipeline and we obtain the correlated flux in six FT spectral channels (ranging from $1.99 - 2.38\, \upmu$m) and in medium resolution for the SC (233 wavelength bins in the range $1.97 - 2.48\,\upmu$m). With the chosen setup, where the source position is varied on only one of the two beams forming a baseline, the measured correlated flux is given by
\begin{equation}
\av{\eta_i^\ps\left(\vec{\alpha}_0\right) \left(\eta_j^\ps\left(\vec{0}\right)\right)^*}_\obj = 
A_i\left(\vec{\alpha}_0\right) e^{i\phi_i\left(\vec{\alpha}_0\right)}\, 
A_j\left(\vec{0}\right) e^{-i\phi_j\left(\vec{0}\right)}\,.
\end{equation}
Thus, the measurement directly scans the phase and amplitude maps on the modulated channel. Potential offsets in the accompanying non-modulated beam, $\phi_j\left(\vec{0}\right)\neq 0$, can only cause a global phase shift, which we fit and remove in the subsequent analysis. Finally, we consider the amplitude maps normalized to their maximum value, such that $A_\jj\left(\vec{0}\right)$ has no impact on our result. 

\begin{figure}
\includegraphics[]{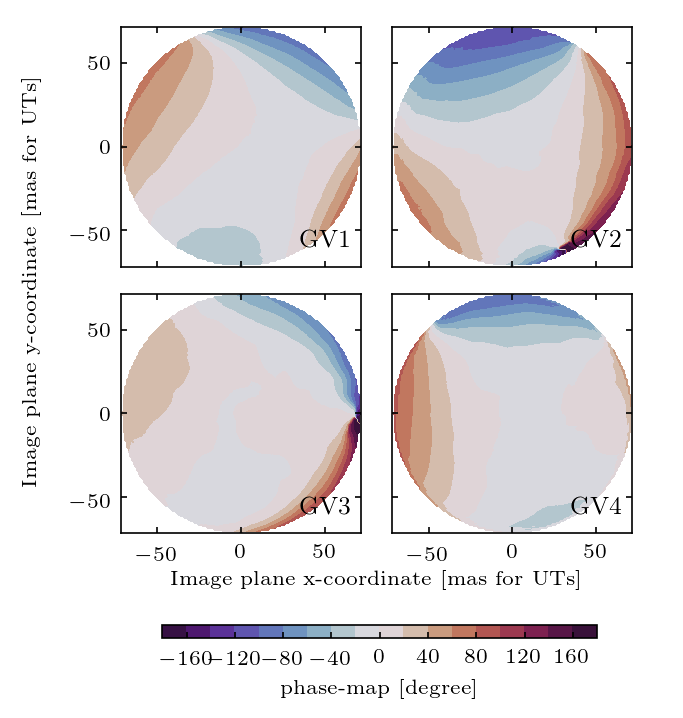}
\caption{Science channel phase maps reconstructed by the procedure of Sec. \ref{sec: measurements} from the Calibration Unit measurement on 03/03/20 for all four GRAVITY beams.}
\label{fig: measurement-phasemap}
\end{figure}

In summary we apply the following analysis steps to obtain the FT and the differential SC-FT phase and amplitude maps.
\begin{enumerate}
\item We fit and subtract a linear time drift from the phases measured in each spectral channel and on each baseline.
\item Phases and amplitudes are binned on a spatial grid with resolution $1\,\mathrm{mas}$ and averaged over all periods of in- and out-spiral available.
\item The image plane coordinates do not align perfectly with the amplitude maximum, i.e. the source position for which the coupling to the fiber is most efficient. We correct for this effect by fitting a Gaussian profile and shifting the coordinate origin to its maximum. 
\item Interpolation over the gridded data gives one phase and amplitude map per spectral channel and baseline.
\item All spectral channels are combined into a single map at reference wavelength $\lambda_0 = 2.2\,\mu\mathrm{m}$, by applying the approximate coordinate scaling from Eq. (\ref{eq: overlap-approximate-scaling}). Here, we verified that the individual maps are consistent over the full spectral range. Cross-validation with simulated maps shows that the error introduced by the approximate scaling relation is small, apart from the very margins of the map. It further cancels between channels above and below $\lambda_0$ to a very good degree.
\item From consideration of all baselines, three maps are available for each telescope. We again verify their consistency and average them into a single phase and amplitude map.
\end{enumerate}
This method results in a FT and a differential SC-FT map for each telescope. Subtracting the former from the latter, we finally arrive at the desired SC phase map, which is shown in Fig.~\ref{fig: measurement-phasemap}. The amplitude map on the SC, on the other hand, is measured directly.

The Calibration Unit measurement was performed twice with a four month break, in late-2019 and early-2020, and we use the data to construct two independent sets of maps. These agree very well in the qualitative features and structures displayed. On the quantitative level the maps display moderate differences of the order of $\sim 10^\circ$, which are smaller at the center and increase towards the map's margins.
\subsection{Representation in the pupil plane}
\label{sec: zernike}
Analyzing the Calibration Unit measurement as described in the previous subsection, we obtain the phase and amplitude maps on a grid discretizing the image plane. We use this result to infer the underlying pupil-plane and fiber aberrations, $d_\pup\left(\vec{u}\right)$ and $d_\foc\left(\vec{u}\right)$, in their Zernike representation. To this end, we developed a simulation tool that creates complex maps of image-plane distortions from a set of Zernike coefficients according to Eq.~(\ref{eq: zernike-decomposition}), Eq.~(\ref{eq: overlap-complex-pupil-function}) and Eq.~(\ref{eq: overlap-pointsource}).

For the fit we consider the two Calibration Unit measurements from 2019 and 2020 separately and combine the phase and amplitude maps for each telescope into a complex map. We then minimize the square absolute difference to the model prediction summed over all pixels with respect to the input coefficients. Due to the nature of the approximate coordinate scaling (step 5 of the analysis pipeline), at a map's edge only the smallest wavelengths contribute. We limit the radius to which the data is considered in the fit to $\alpha_\ma\times\lambda_\mathrm{low}/\lambda_\mathrm{high}$. With $\alpha_\ma$ being the size of the full map and $\lambda_\mathrm{low}$ and $\lambda_\mathrm{high}$ the wavelength of the lowest and highest channel, respectively. This choice ensures equal participation of all channels in the fit.

The optical layout of observations with the Calibration Unit has some important differences with the on-sky situation, for which the phase maps will be applied later. Namely, the lack of a central obscuration and an enlarged outer stop $r_\mathrm{GCU} = 9.6\,\mathrm{m}/2$ alter the shape of the pupil defined in Eq.~(\ref{eq: overlap-ppupil}). As a consequence, the Calibration Unit pupil illuminates image-plane aberrations out to a slightly larger radius. We choose to normalize the Zernike polynomials by $r_\tel = 8.0/2\,\mathrm{m}$, i.e. the telescope area covered by the secondary mirror, to optimize our parameterization for the on-sky case. Image plane distortions, on the other hand, are normalized over the image-plane fiber width at $\lambda_0$, $\tilde{\sigma}_\fib = \epsilon\lambda_0/\left(4r_\tel \sqrt{\ln 2}\right)$, i.e.
\begin{equation}
d_\foc \left(\vec{\alpha}\right) = \sum_{n=0}^{n_\ma} \sum_{m=-n}^{m} B_n^m\, Z_n^m\left(\vec{\alpha}/\tilde{\sigma}_\fib\right)\,.
\label{eq: zernike-fits-focal-plane-decomposition}
\end{equation}

\begin{figure}
\includegraphics[]{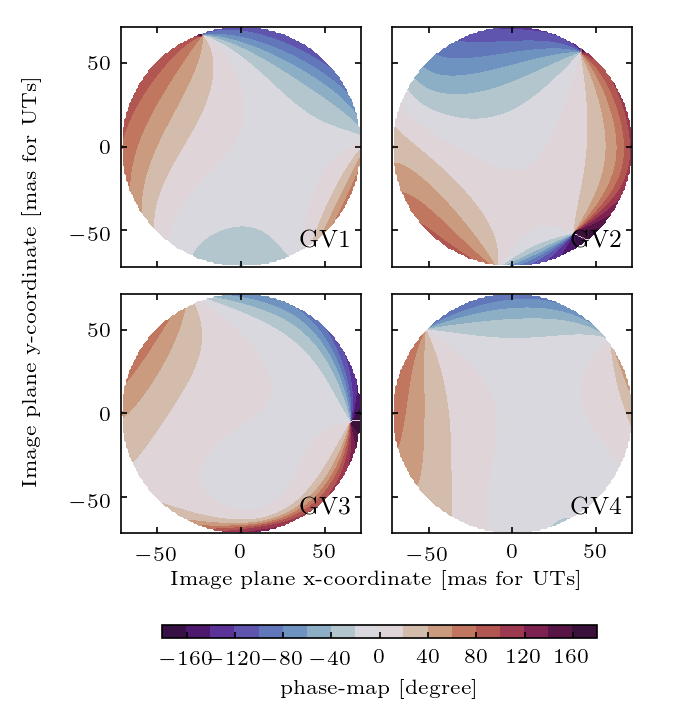}
\caption{Science channel phase maps obtained from fits to the differential SC-FT maps, measured on 03/03/20 for all four GRAVITY beams.}
\label{fig: phase-maps-fits}
\end{figure}
\begin{figure}
\includegraphics[]{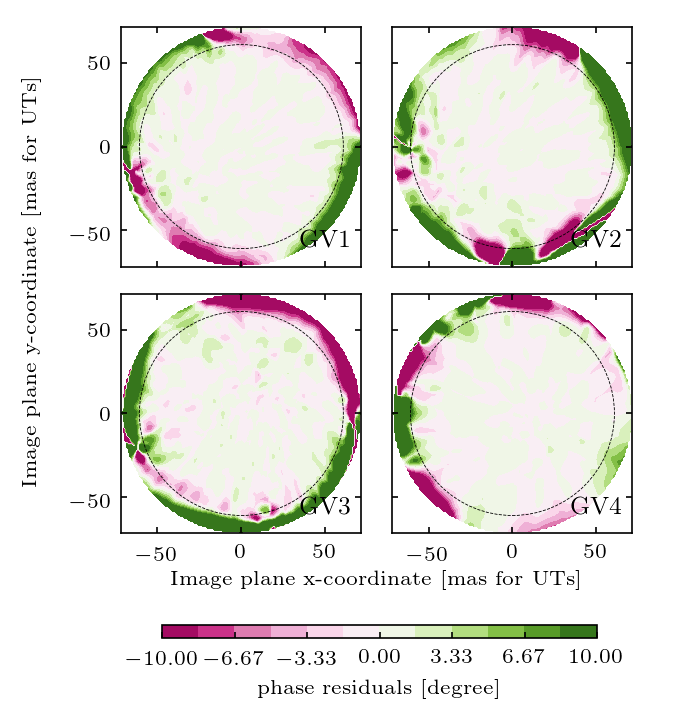}
\caption{Phase residuals of the fit to the differential SC-FT map measured on 03/03/20 for all four GRAVITY beams. Only the data within the dashed circle is considered in the fit; at larger radii the cancellation of wavelength-dependent scaling errors is not guaranteed.}
\label{fig: phase-maps-residuals}
\end{figure}
Of the different types of maps constructed, the fringe tracker provides the cleanest system and thus gives an important benchmark point for the agreement between model and data. We thus use the FT-maps to determine the order $n_\ma$ to which Zernike polynomials in the pupil- and focal-plane aberrations are considered. Successively increasing the fit order, we find that pupil-plane aberrations with $n_\ma = 6$ and focal-plane aberrations with $n_\ma=2$ provide satisfactory model consistency, while still allowing for manageable convergence times. Increasing the Zernike order in the pupil plane is especially important to reduce phase residuals at larger radii, while the central part of the maps can also be described by polynomials of lower order. Fits without focal-plane aberrations manage to reproduce the phase structure to a satisfactory degree, but show poor consistency between the phase and the amplitude data. Finally, an additional parameter accounts for the overall amplitude scaling between measured and predicted maps, such that each fit constrains at least 34 degrees of freedom. The phase RMS achieved for the fringe tracker fits is of order $\sim 1^\circ$ for all beams and data sets; extrapolation of the fit result to the full map radius yields an RMS of a few degrees.

In principle, it is possible to directly fit the SC maps by the same procedure employed for the FT. However, by further refining the analysis we can remove additional systematic effects from the SC maps. Creating the maps, we corrected for misalignment of the image plane coordinates with the amplitude maximum (step 3 in the analysis pipeline). This shift, however, is not guaranteed to be identical on SC and FT, and as a result there can be a small offset between the FT phase entering the differential SC-FT measurement. To describe this effect, we fit a differential map, predicted from two sets of Zernike coefficients, to the SC-FT maps. The latter of this two sets of parameters is largely fixed to the previously obtained FT coefficients, and only the tip-tilt terms are allowed to vary. The SC parameters, on the other hand, are all free, such that the fit eventually determines the desired SC maps and the offset between the two channels. 

From the best-fit coefficients of the differential SC-FT fit, which we summarize in App. \ref{sec: zernike-coefficients}, we reconstruct a complex SC map. Its phase is displayed in Fig.~\ref{fig: phase-maps-fits}. As expected, the structure agrees very well with the maps obtained by direct evaluation of the Calibration Unit measurement in Fig.~\ref{fig: measurement-phasemap}. Residuals between measured and fitted SC-FT map, shown in Fig.~\ref{fig: phase-maps-residuals}, are low over the full radius considered for the fit. We obtain a best-fit RMS of $1^\circ - 2^\circ$ for most beams and data sets and two slightly worse results with RMS $\sim 3^\circ$ and $\sim 5^\circ$. Going to larger radii, the disagreement between fit and data starts to increase. This can be caused either by wavelength-dependent errors or by higher-order aberrations, beyond those considered for the fit. Indeed, in optimizing $n_\ma$, we noted that every increase improved the extrapolation to large separations. However, at such large off-axis distances, fiber damping becomes very significant, resulting in a poor signal-to-noise ratio. Thus, we consider the Zernike decomposition up to 6th order sufficient for our applications.

\section{Application to GRAVITY observations}
\label{sec: application}
Static, field-dependent aberrations affect the visibility measurement whenever the size of an observed object is comparable to the fiber FOV. Here, we apply the formalism developed in Sec.~\ref{sec: formalism} alongside the characterization of aberrations from Sec.~\ref{sec: characterization} to observations of two different binary systems. First, as a proof of concept, we consider a test-case binary observed with the Auxiliary Telescopes (ATs), where the system's position in the FOV was systematically varied and thus screened over the phase and amplitude maps. Second, we apply the aberration-correction to GC observations with the UTs from 2017 and 2018. During those epochs, close to pericenter passage, S2 and Sgr A* where observed simultaneously in a single fiber pointing.

The data considered in either analysis consists of visibility amplitudes, squared visibilities and closure phases with a relative weighting of (1:1:2). To infer the sources' separation, we fit a binary model based on Eq.~(\ref{eq: visibility-observed-binary}), which we extend to account for the effect of finite spectral resolution and for a homogeneous background with flux ratio $f^\bkg$ relative to the first binary component,
\newpage
\begin{strip}
\begin{equation}
V^{\obs}_\bin \left(\vec{b}_{i,j}/\lambda\right)
=\frac{
\tilde{A}_i\left(\vec{\alpha}_1\right) \tilde{A}_j\left(\vec{\alpha}_1\right)  V_\lambda\left[ \left(\vec{b}_{i,j}\cdot\vec{\alpha}_1 - \vec{d}_{i,j}\left(\vec{\alpha}_1\right) \right), \nu_1\right]
+
\tilde{A}_i\left(\vec{\alpha}_2\right) \tilde{A}_j\left(\vec{\alpha}_2\right)  V_\lambda\left[ \left(\vec{b}_{i,j}\cdot\vec{\alpha}_2 - \vec{d}_{i,j}\left(\vec{\alpha}_2\right) \right), \nu_2\right]
}
{\sqrt{\prod_{x=i,j}
\left[\tilde{L}_x\left(\vec{\alpha}_1\right) V_\lambda\left(\vec{0}, \nu_1\right) + f^\bin \tilde{L}_x\left(\vec{\alpha}_2\right)  V_\lambda\left(\vec{0}, \nu_2\right) + f^\bkg V_\lambda\left(\vec{0}, \nu_\bkg\right) \right]}} \,.
\label{eq: binary-fit}
\end{equation}
\end{strip}%
\noindent Phase distortions enter this expression via the OPD correction $d_{i,j} = \left(\tilde{\phi}_i-\tilde{\phi}_j\right)\times\lambda/2\pi$. Further, the point-source visibility averaged over a spectral channel is
\begin{equation}
V_\lambda\left(\vec{d}, \nu\right) = \int d\lambda\, P\left(\lambda\right) \left(\frac{\lambda}{2.2\, \upmu m}\right)^{-1-\nu} e^{-2\pi i\, d/\lambda}\,.
\end{equation}
The spectral bandpass $P\left(\lambda\right)$ is given by a top hat function. The source positions $\vec{\alpha}_1$ and $\vec{\alpha}_2$, the flux ratios $f^\bin$ and $f^\bkg$ as well as the spectral index of the central component ($\nu_1$) and the background flux ($\nu_\bkg$) are free fit parameters, while the companion's spectral slope is fixed to $\nu_2=3$.

Finally, $\tilde{A}_{i/j}$, $\tilde{\phi}_{i/j}$ and $\tilde{L}_{i/j}$ in Eq.~(\ref{eq: binary-fit}) refer to the phase maps, amplitude maps and the photometric lobes as they are encountered in on-sky observations. Those have two important differences with the Calibration Unit measurement. Firstly, while the pupil-plane representation of the aberrations is the same for both settings, the presence of a central obscuration and the smaller outer stop affects the realization of the maps in the image plane. This is conveniently captured by using the Zernike coefficients found in Sec.~\ref{sec: zernike} to create a new set of maps with adjusted pupil configuration. Secondly, the maps are subject to turbulent smoothing according to Eqs.~(\ref{eq: smoothing-photometric-lobe}) and (\ref{eq: smoothing-photometric-lobe}).
\subsection{Verification for a binary test-case}
\label{sec: binary-test-case}
\begin{figure}
\includegraphics[width=\linewidth]{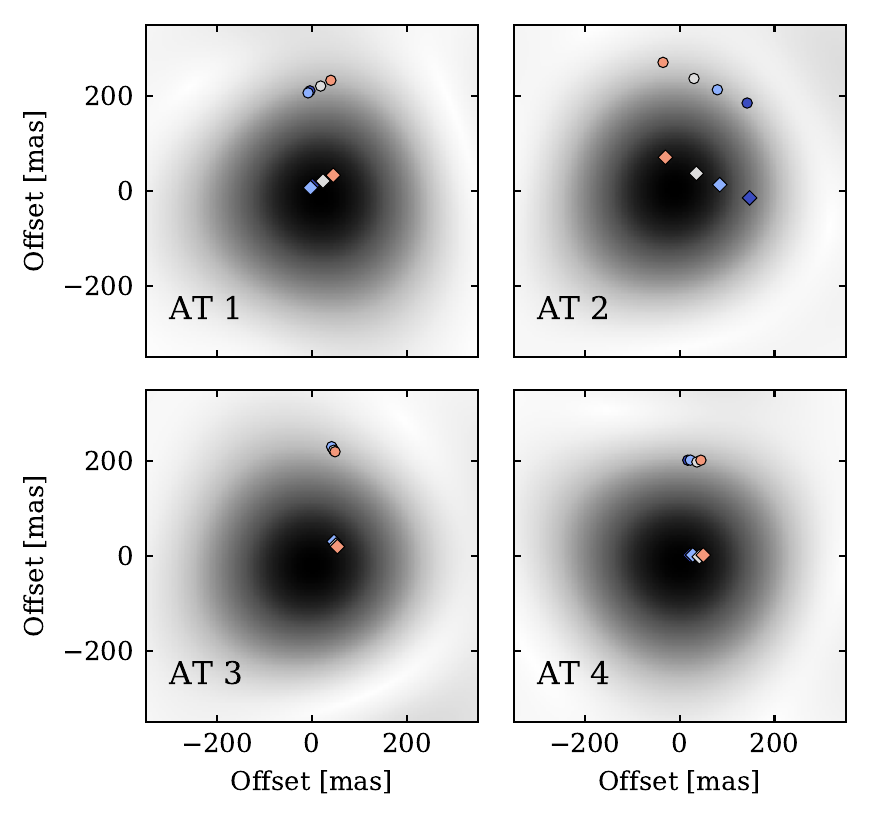}
\caption{Illustration of the AT binary test observations, showing the position of the two binary components (circles and diamonds, respectively) relative to the fiber profile (gray shading). Color gradients are chosen in accordance with Fig.~\ref{fig: binary-result}. For this test, the fiber position was varied on AT2 only, but kept fixed on the other three telescopes.}
\label{fig: binary-setup}
\end{figure}

The test-case observations, carried out with the ATs in astrometric configuration, targeted HIP 41426, a binary with K-band magnitude $m_\mathrm{K}\simeq 5.393$ at $\RA=8{:}26{:}57.75\,\mathrm{h}$, $\DEC=-52{:}42{:}17.8$ \citep{2003yCat.2246....0C}. The system has an approximate separation of 200 mas. Its position relative to the GRAVITY fiber was kept fixed for three of the four telescopes and varied in 24 steps between $\pm 400\, \mas$ on AT2. At each offset, ten frames with a 6 s integration time were taken. The setup is illustrated in Fig.~\ref{fig: binary-setup}, which shows both binary components relative to the fiber profile on all four telescopes. The shift was applied along the x-axis in the frame of the GRAVITY pupil, whose rotation with respect to the field results in a diagonal movement on the sky.

We use the Zernike coefficients obtained for the SC in Sec.~\ref{sec: zernike} to produce phase and amplitude maps tailored to observations with the ATs. In this case, the pupil, c.f. Eq.~(\ref{eq: overlap-ppupil}), is defined by $r_\tel = 1.82\,\mathrm{m}/2$ and $r_\central = 0.14\,\mathrm{m}/2$. After beam collimation, ATs and UTs illuminate the same section on the GRAVITY mirrors, such that the pupil-plane phase screen can simply be scaled to the AT radius, i.e. $r_\tel = 1.82\,\mathrm{m}/2$ also applies in the Zernike decomposition of Eq.~(\ref{eq: zernike-decomposition}). To authenticate the impact of correct aberration modeling, we compare our results to a second, no-map analysis. In this latter scenario, we set all phase maps to zero and all amplitude maps to one, i.e. $\tilde{\phi}_{i/j}=0$, $\tilde{A}_{i/j}=1$. 

\begin{figure}
\includegraphics[width=\linewidth]{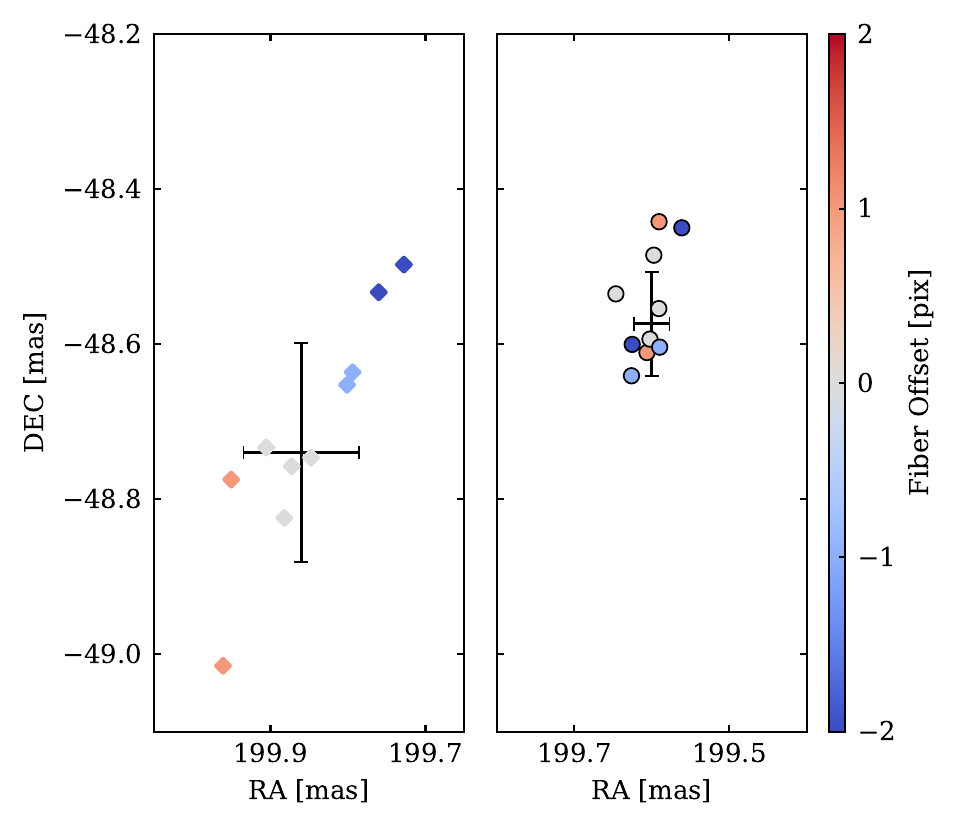}
\caption{Binary separation inferred for a varying fiber offset on AT2 with (right panel) and without (left panel) application of the phase and amplitude maps. Each data point shows the average over two polarization states, and the range of offsets corresponds to $\pm 200\,\mas$, approximately.}
\label{fig: binary-result}
\end{figure}

For too large fiber offsets, the signal-to-noise ratio on AT2 is poor due to large fiber damping and we consequently discard these data. The remaining pointings are shown in Fig.~\ref{fig: binary-setup}, and the corresponding separation, measured from a binary fit to the data according to Eq.~(\ref{eq: binary-fit}), is given in Fig.~\ref{fig: binary-result}. 

The AT binary test-case clearly validates our aberration corrections. Different configurations yield consistent results only if phase and amplitude maps are considered in the analysis. Including the correct aberration model in the analysis clearly shifts the result and reduces the scatter. Even more importantly, however, the separation found in the no-map analysis systematically depends on the fiber position; it is largest for positive fiber-offsets and smallest for offsets in the negative direction. With application of the aberration-correction, this systematic is largely removed. 

We consider the binary test-case observations primarily as a proof of concept and therefore forgo a full analysis of the measurement's systematic error as carried out for the GC. Such uncertainties arise from the accuracy to which the phase maps can be determined and from the uncertainty of the atmospheric smoothing kernel. Further, there can be minor differences in the phase and amplitude maps between AT und UT observations, and our treatment is optimized to the UT scenario.

As the shift in its central value indicates, the binary separation is large enough that even at perfect fiber pointing at least one source lies in a region of the FOV where aberration-induced phase errors are significant. Accurate astrometry thus is not a question of precise fiber alignment but is only possible with a consistent treatment of the pupil-plane distortions in the analysis. 

\subsection{The separation between S2 and Sgr~A*}
\label{sec: s2-corrections}
\begin{figure}
\includegraphics[]{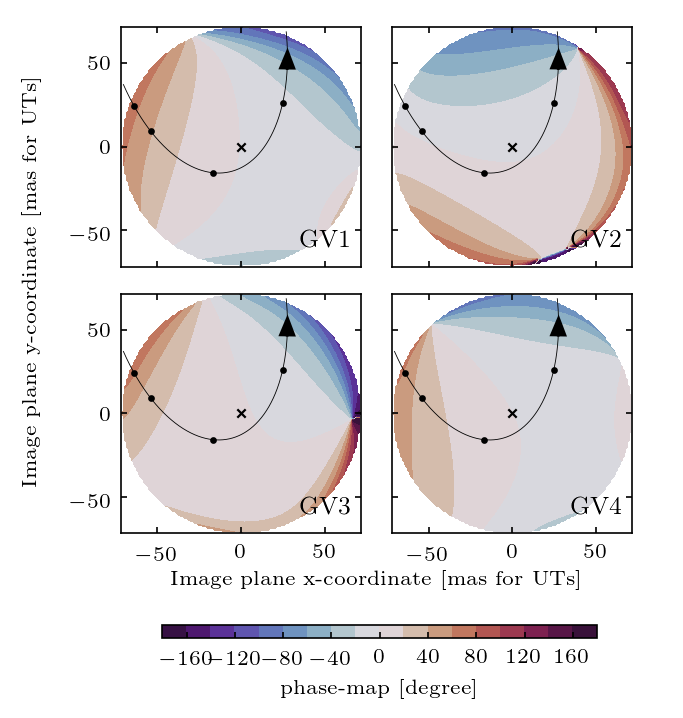}
\caption{The orbit of S2 relative to the phase maps as applied for the GC analysis (measurement from 03/03/20, $\sigma_t=10\,\mathrm{mas}$). Dots indicate the position of S2 on 2017.2, 2017.6, 2018.2 and 2018.7, respectively, while the cross marks Sgr A*.}
\label{fig: phase-maps-gc}
\end{figure}

Having verified our approach to correct for aberration-induced systematic errors, we also apply it to Galactic Center observations with GRAVITY. During 2017 and 2018, i.e. close to pericenter passage, S2 and Sgr~A* where observed simultaneously in a single fiber pointing. In particular during 2017, when the off-axis distance of S2 was larger, the aberration correction improves the inferred binary separation. In 2019, in contrast, the Sgr~A*-S2 separation exceeds the single telescope beam size of about $60\,\mas$, and GRAVITY observes both sources separately in so called dual-beam mode. Their separation is then obtained by calibrating Sgr~A* with S2 and fitting a point source model to its visibilities (see \cite{2020A&A...636L...5G} for details). In this configuration, each source can be well aligned with the fiber center, such that field-dependent aberrations do not impact the measurement.

To derive the aberration-induced shift of the S2 position, we examine a subset of the GRAVITY data used in \cite{2019A&A...625L..10G}. In particular, we apply stricter quality cuts and demand a high signal-to-noise ratio. Phase and amplitude maps are generated from the coefficients obtained in Sec.~\ref{sec: zernike} by accounting for the specific geometry of UT-observations, i.e. $r_\tel = 8.0\,\mathrm{m}/2$ and $r_\central = 0.96\,\mathrm{m}/2$. The residual turbulent tip-tilt is between $10\,\mathrm{mas}$ and $15\,\mathrm{mas}$ per axis \citep{2019A&A...625A..48P}. In total, we consider four different realizations of the aberration maps which are given by the independent analysis of the two calibration measurements in 2019 and 2020 each convolved with the minimum and maximum smoothing assumption. A representative example for the phase maps applied in the GC analysis is shown in Fig.~\ref{fig: phase-maps-gc} in relation to the orbit of S2.

\begin{figure*}
\begin{center}
\includegraphics[width=0.7\linewidth]{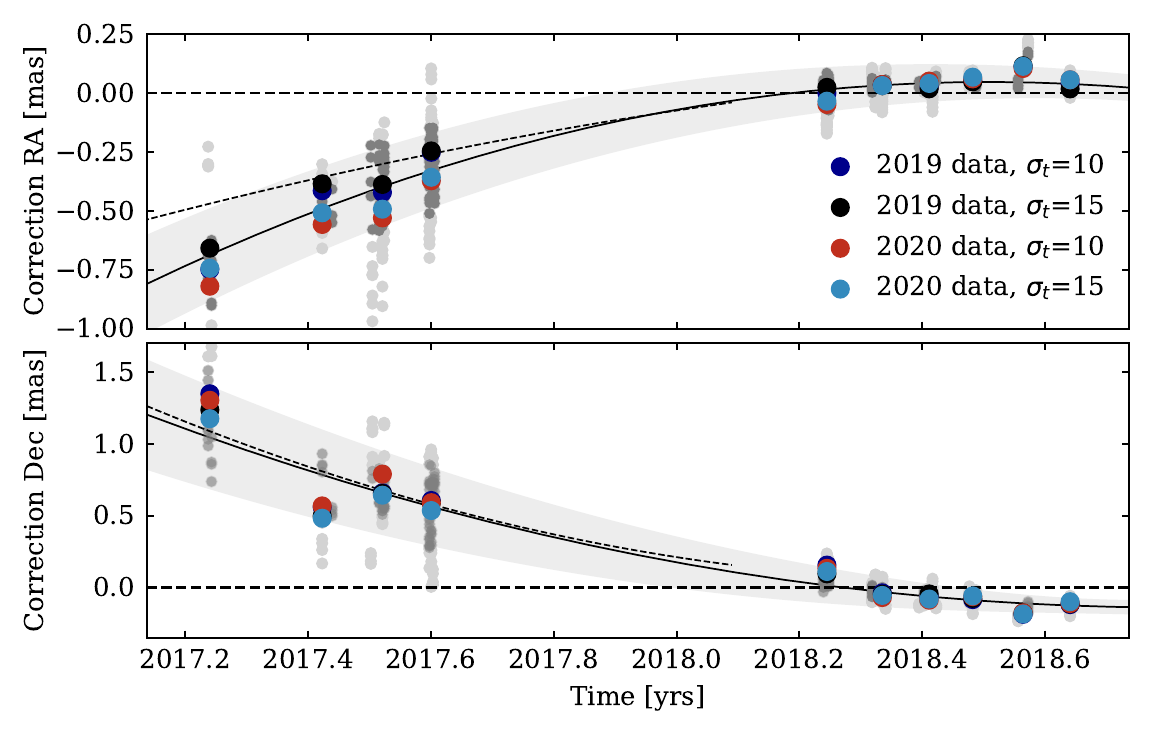}
\caption{The difference in S2 position obtained from an analysis with and without application of the aberration corrections. Colored dots indicate the epoch-wise mean for different realizations of the phase and amplitude maps, gray dots the results for individual observations. From these, we determine a mean position-correction as function of time with a corresponding upper and lower limit as indicated by the black solid line and the gray band. The thin dashed line, finally, represents the correction applied in \cite{2019A&A...625L..10G}.}
\label{fig: gc-results}
\end{center}
\end{figure*}

Our main result, the difference in S2 position with and without aberration-corrections averaged per month, is shown in Fig.~\ref{fig: gc-results}. As expected, the correction is largest in early-2017 and smallest around peri-center passage in May 2018. Further, the mean corrections per epoch obtained with the four different realizations of the aberration maps are consistent over the full observational period.

As the orbit of S2 smoothly scans over the phase and amplitude maps (see Fig.~\ref{fig: phase-maps-gc}), we also expect a smooth variation in the position-correction. Indeed, the time-dependence in Fig.~\ref{fig: gc-results} is well described by a second-order polynomial fit
\begin{align}
\Delta\RA &= \left(-0.44\, \tau^2 + 0.11\, \tau + 0.04\right)\,\mas \,,\\
\Delta\DEC &= \left(0.41\,\tau^2 - 0.47\,\tau - 0.06\right)\,\mas\,,
\label{eq: s2-correction}
\end{align}
where $\tau = t/\mathrm{years}-2018.4$ refers to the shifted observation date in years.

In addition to the mean correction per epoch, Fig.~\ref{fig: gc-results} also shows the individual file-by-file results as gray dots. These give some insight into the uncertainty of the aberration-correction. When we fit the orbit of S2, any such uncertainty must to be propagated as source of systematic error. We construct a upper and a lower estimate of the correction, containing 67\% of the files per epoch. This is shown in Fig.~\ref{fig: gc-results} as a gray band.

Apart from the systematic error, we also need to account for the statistical uncertainty of the S2 position. That is, as the phase and amplitude error changes when the S2 position is varied within its errorbars, we need to propagate this effect to the final correction. To this end, we take the position error of the original, un-corrected data point from which we draw 100 realizations and shift the aberration maps by it. We then derive the correction from each realization independently and use their scatter to estimate the statistical error of the S2 position correction. The resulting mean statistical uncertainty per epoch is small, between $10\,\mu\mathrm{as}$ and $30\,\mu\mathrm{as}$, but we nevertheless also account for it in the orbit fitting.

A further check is to ask the question, what correction makes the 2017 and 2018 GRAVITY positions optimally match to the rest of the S2 data. To this end, we included a scaling factor $f_\mathrm{corr}$ in the correction we apply, such that $f_\mathrm{corr}=1$ is our best correction and $f_\mathrm{corr}=0$ is no correction. This parameter we can then include in the orbit fit (see Sec.~\ref{sec: gc-distance}). The best fit yields 
 $f_\mathrm{corr}=0.99 \pm 0.06$, i.e. identical to the correction we have derived purely from calibration data. This gives an independent confirmation of our concept and the resulting aberration correction: Our correction yields the most consistent S2 orbit.
  
The aberration correction presented here constitutes a further refinement of the analysis in \cite{2020A&A...636L...5G}. There, we applied the measured aberration maps as shown in Fig.~(\ref{fig: measurement-phasemap}) directly, rather than the fitted decomposition in terms of pupil-plane Zernike polynomials. To account for the widening of the maps, which occurs when projecting from the enlarged stop on the Calibration Unit to the telescope pupil, in addition to the effect of turbulence, we applied a smoothing kernel of $\sigma_t = \left( 19 \pm 5 \right)\,\mathrm{mas}$. The resulting best-estimate for the correction is depicted in Fig.~\ref{fig: gc-results} as dashed line. Both methods give consistent results, affirming the robustness of the approach. The only sizable deviation is in 2017.2, when S2 was observed at a separation comparable to the maximum radius for which we obtained the calibration measurement (see Fig.~\ref{fig: phase-maps-gc}). This case shows the strength of the Zernike decomposition, which allows for a well-defined extrapolation.

\section{Results}
\label{sec: results}
\subsection{Determination of the S2 orbit}
\begin{table}
\small
\renewcommand{\arraystretch}{1.5}
\begin{tabular}{l|c}
misalignment between mass and IR-emission & $12\,\pc$\\
\hline
wavelength calibration of SINFONI & $9\,\pc$\\
\hline
GRAVITY astrometry & $29\,\pc$ \\
\hspace{.7cm}baseline accuracy &\hspace{.7cm} $4\,\pc$ \\
\hspace{.7cm}wavelength accuracy &\hspace{.7cm} $9\,\pc$ \\
\hspace{.7cm}model \& data selection &\hspace{.7cm} $9\,\pc$ \\
\hspace{.7cm}atmospheric differential dispersion &\hspace{.7cm} $5\,\pc$ \\
\hspace{.7cm}aberration-correction &\hspace{.7cm} $23\,\pc$\\
\hspace{.7cm}metrology correction &\hspace{.7cm} $10\,\pc$ \vspace{0.3cm}
\end{tabular}
\caption{Contribution to the systematic errors affecting the measurement of $R_0$, for details see \cite{2019A&A...625L..10G}. Adding all contributions quadratically, we find a total systematic uncertainty of $33\,\pc$.}
\label{tab: discussion-systematics}
\end{table}

In the following we evaluate the effect of the aberration correction on the S2 orbit. The data used is similar to \cite{2020A&A...636L...5G} and described in detail in Appendix \ref{sec: gc-data}. We employ the same fitting procedure as in \cite{2020A&A...636L...5G}, using a 13-parameter, Post-Newtonian orbit model. Six of those parameters describe the Kepler orbit ($a$, $e$, $i$, $\omega$, $\Omega$, $t_\mathrm{peri}$), and another six describe the reference frame relative to the AO spectroscopy and assumed Local Standard of Rest (LSR) correction, ($x_0$, $y_0$, $R_0$, $\dot{x}_0$,  $\dot{y}_0$, $\dot{z}_0$). Here, $R_0$ is the distance to the GC, the prime focus of this work, and $M_\bullet$ the central mass. The best-fit parameters are given in Tab.~\ref{tab: discussion-fitresults}.

For determining the systematic uncertainty, we follow the approach in  \cite{2019A&A...625L..10G} of varying our assumptions and tracing the associated changes in $R_0$. Compared to our earlier work, we also include the uncertainty due to the aberration correction, as given by the gray band in Fig.~\ref{fig: gc-results}. The individual contributions are given in Tab.~\ref{tab: discussion-systematics}. It turns out that the aberration correction is the dominant contributor to the systematic error. The total systematic uncertainty is $33\,$pc when adding the contributions quadratically. 

Our best estimate of the Galactic Center distance thus is 
\begin{equation}
R_0 = 8275 \pm \left.9\right|_\estat \pm \left.33\right|_\esys ~\mathrm{pc}\,.
\end{equation}

\begin{table}
\small
\begin{center}
\renewcommand{\arraystretch}{1.3}
\begin{tabular}{c|c}
parameter & value
\\
\hline
$a~[\mas]$  &
$124.982 \pm 0.034$ 
\\
$e$ & 
$0.884215 \pm 0.000058$
\\
$i~[\deg]$ & 
$134.685 \pm 0.029$
\\
$\omega~[\deg]$ & 
$66.259 \pm 0.030$ 
\\
$\Omega~[\deg]$ & 
$227.175 \pm 0.029$  
\\
$P~\left[\mathrm{yr}\right]$ & 
$16.0458 \pm 0.0013$ 
\\
$t_\mathrm{peri}~\left[\mathrm{yr}\right]$ &
$2018.378990 \pm 0.000082$ 
\\
$x_0~\left[\mathrm{mas}\right]$ & 
$-0.79\pm 0.10$ 
\\
$y_0~\left[\mathrm{mas}\right]$ & 
$0.00 \pm 0.11$ 
\\
$\dot{x}_0~\left[\mathrm{mas/yr}\right]$ & 
$0.0780 \pm 0.0091$ 
\\
$\dot{y}_0~\left[\mathrm{mas/yr}\right]$ & 
$0.0342 \pm 0.0094$ 
\\
$\dot{z}_0~\left[\mathrm{mas/yr}\right]$ & 
$-2.6 \pm 1.4$ 
\\
$M_\bullet~\left[10^6\,M_\odot\right]$ & 
$4.297 \pm 0.013$ 
\\
$R_0~\left[\mathrm{pc}\right]$ &
$8274.9 \pm 9.3$ 
\end{tabular}
\end{center}

\caption{Orbital parameters of S2 with their statistical uncertainties.}
\label{tab: discussion-fitresults}
\end{table}
\label{sec: gc-distance}
\subsection{Comparison to previous results}
\begin{figure*}
\begin{center}
\includegraphics[width=0.7\linewidth]{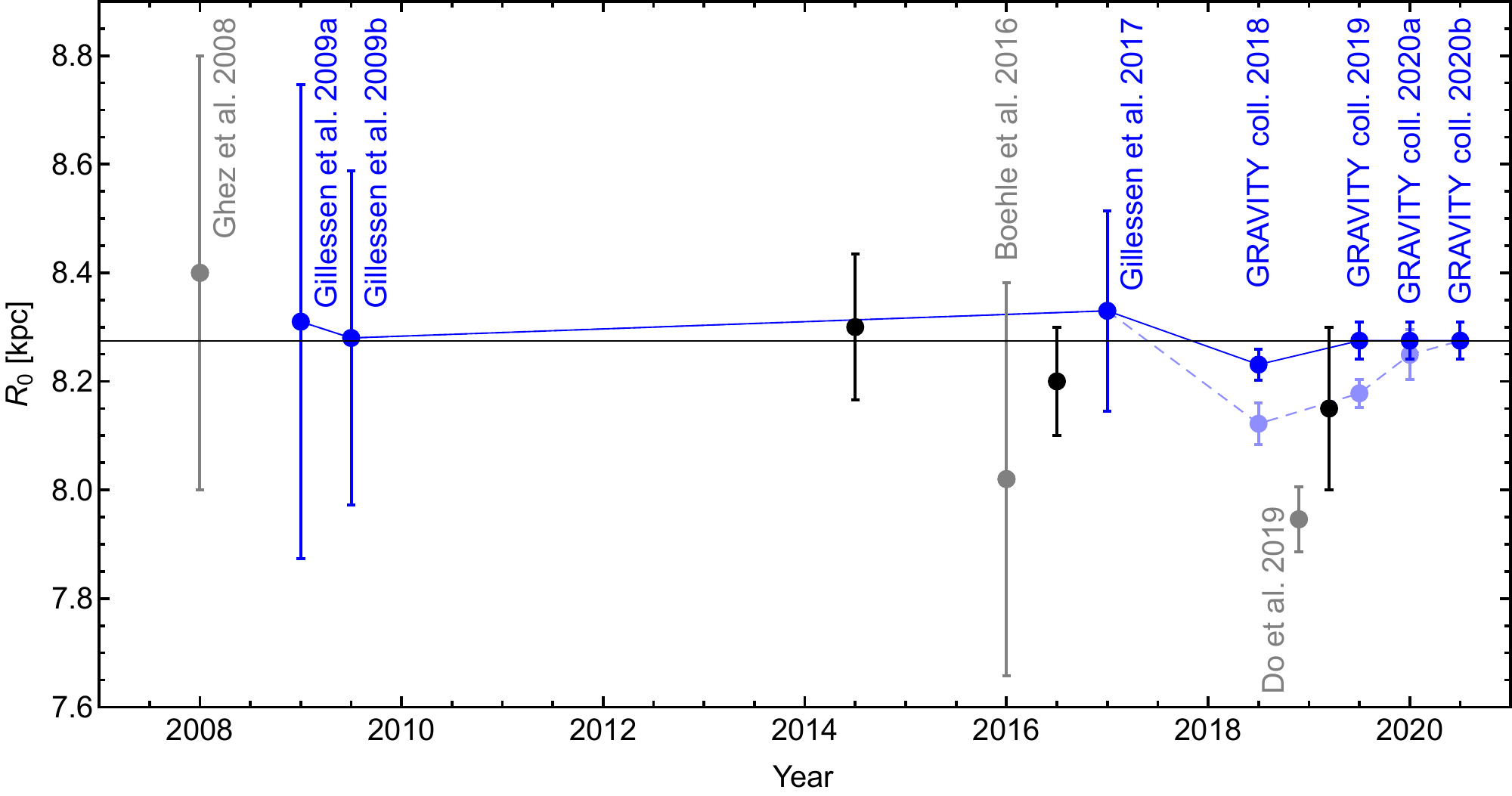}
\caption{Measurements of the Galactic Center distance over time with a focus on studies of the S2 orbit. Blue points show results obtained with the SINFONI, NACO and GRAVITY data with (dark blue) and without (light blue) application of the aberration corrections. Gray $R_0$ determinations are based on data from the Keck observatory. For comparison, we show in black results based on the statistical parallax of the nuclear star cluster \citep{2015MNRAS.447..948C} and from modeling the Milky Way dynamics based on observations of molecular masers \citep{2019ApJ...885..131R}. \cite{2016ARA&A..54..529B}, finally, give the GC distance based on a combination of various methods.} 
\label{fig: r0-fig}
\end{center}
\end{figure*}
\nocite{2008ApJ...689.1044G, 2009ApJ...692.1075G, 2016ApJ...830...17B}

\begin{table}
\tiny
\renewcommand{\arraystretch}{1.3}
\begin{tabular}{l|ccc}
Phasemaps & None & 2017 only & 2017 and 2018 \\
\hline
GRAV. coll 2018&${\bf 8122 \pm 31}$&&$8231 \pm 16 \pm 24$\\
GRAV. coll. 2019&${\bf 8178 \pm 13 \pm 22}$ && $8275 \pm 13 \pm31$\\
GRAV. coll. 2020&&${\bf 8249 \pm 9 \pm 45}$ & $8275 \pm 9 \pm 33$\\
  this work  &&$8246 \pm 9 \pm 33$ & ${\bf 8275 \pm 9 \pm 33}$
\end{tabular}

\caption{Published values of $R_0$ {\bf (bold)} and the corresponding values if the aberrations are taken into account (right column). All values in pc.}
\label{tab: r0-history}
\end{table}

Our previous determinations of the GC distance in \cite{2018A&A...615L..15G}, \cite{2019A&A...625L..10G} and \cite{2020A&A...636L...5G} were biased by the field-dependent aberrations. Taking them into account brings all our measurements into agreement as shown in Fig.~\ref{fig: r0-fig} and Tab.~\ref{tab: r0-history}. We further note the following:

\begin{itemize}
\item In contrast to \cite{2020A&A...636L...5G}, we also apply a correction for the 2018 data, where S2 and Sgr~A* were close to each other and close to the field center. Yet, the small aberration corrections lead to a small upward correction of $R_0$ of around $30\,$pc, comparable to the systematic error.
\item The orbit is particularly sensitive to the pericenter data. This leads to the effect that the statistical uncertainty decreases strongly with time, while the systematic uncertainty even increases slightly during this time frame, since varying the assumptions then leads to stronger variations in the fit result.
\end{itemize}

\label{sec: gc-previous}
\subsection{Comparison with further S2-based results}
\label{sec:AppA6}
We estimate that the accuracy of our VLT-based result is at the $40\,\pc$ level. However, it deviates significantly from the Keck-based value reported in \citet{2019Sci...365..664D}, with the difference being at the $300\,\pc$ level. Since both works use the orbit of S2 around Sgr~A* for the determination of $R_0$, it is important to investigate where the discrepancy is arising, and we address this in App.~\ref{sec:AppA}. Overall, we conclude that the combination of
\begin{itemize}
\item a difference in the radial velocity data and
\item a modest offset of the Keck coordinate system in the declination direction
\end{itemize}
might explain the discrepancy. Both effects contribute roughly 50\%.

About $20\%$ of the radial velocity difference can be attributed to the Doppler formula in StarKit used implicitly by \citet{2019Sci...365..664D}. The remaining 80\% are unexplained and could be in either the Keck or the VLT data.

The origin of the coordinate system offset is unclear as well. Trying to explain the offset with a shift of the VLT coordinate systems is much harder than imposing a shift of the Keck one due to the high precision of the GRAVITY data.

\section{Conclusions}
\label{sec: discussion}
\begin{figure}
\includegraphics[width=\linewidth]{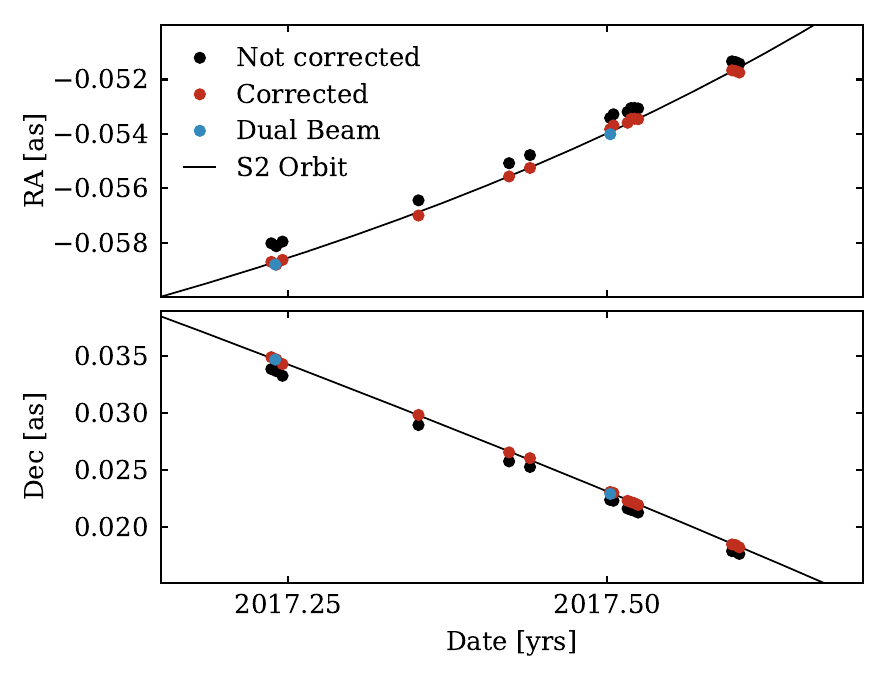}
\caption{Detailed view of the S2 orbit in 2017. Dual-beam points do not suffer from aberration-related systematic errors and agree very well with our corrected data points.}
\label{fig: s2-orbit-2017}
\end{figure}

GRAVITY delivers high-resolution astrometry which, in combination with spectroscopic data, allows for a very precise determination of the Galactic Center distance. The values inferred from different epochs \citep{2018A&A...615L..15G, 2019A&A...625L..10G, 2020A&A...636L...5G} show a small discrepancy at the $1\%$ level, which nevertheless is significant due to the high precision of the measurement. 

We were able to relate this shift to optical aberrations introduced in the instrument, which lead to a field-dependent distortion of the visibility phase. Their effect is the stronger, the further off-axis an object lies within the FOV. In particular Galactic Center observations close to the S2 pericenter passage are affected, where S2 and Sgr A* are detected simultaneously in a single fiber pointing but at a separation comparable to the FOV. In earlier and later epochs, in contrast, we employed the so-called dual-beam method and targeted each source individually. In this case, as for most other GRAVITY science observable, each source can be well centered and aberration corrections become irrelevant. The dual-beam observation mode was also assumed to derive the astrometric error budget in \cite{2014A&A...567A..75L}, which did not include the effect of phase maps for this precise reason.

The full analytical description which we developed here allows us to propagate the effect of optical aberrations at fiber injection to the measured visibilities. Fitting this model to dedicated calibration measurements confirms its validity and enables us to account for the effect in the data analysis. We further verify the approach with dedicated test-case observations.

The formalism which we developed is applicable beyond GRAVITY to any optical/near-IR interferometer where aberrations are introduced in the pupil or the focal plane. There have been several cases in the literature with more than one object lying in the interferometer's FOV, for example some Keck \citep{2013PASP..125.1226C}, CHARA \citep{tenBrummelaar:2005nf} or NPOI \citep{1998ApJ...496..550A} results on binary stars. How severely aberrations affect an observation, however, depends not only on their strength for a particular instrument but also on the off-axis distance considered and on the statistical noise in the measurement. In the example of GRAVITY on the UTs, the mean phase error introduced at $20\,\mas$ separation is $4 - 5$ degrees per telescope and increases to $14 - 20$ degrees at $50\,\mas$. While a binary test case as presented in Sec. \ref{sec: binary-test-case} can serve as a general strategy to diagnose whether aberration-induced systematics are an issue, dedicated calibration measurement are required for their correction in the analysis for each individual instrument.

With the results from the GRAVITY Calibration Unit measurements and our refined analysis scheme, we are able to further improve the separation between S2 and Sgr A* in 2017 and 2018, introducing shifts up to $0.5\,\mas$ caused by the phase aberrations. In Fig.~\ref{fig: s2-orbit-2017}, we show a detailed view of the S2 orbit in 2017, where we have also included two dual-beam measurements that do not suffer from phase aberrations. Indeed, the improved data agrees very well with these positions.

Of all orbital parameters, the distance to the Galactic Center $R_0$ is most strongly affected by the change in the S2 position. This can be easily understood if one views $R_0$ as the scaling factor between angular and proper velocity. As such, the field-dependent phase errors discussed in this work fully explain the shift between earlier $R_0$ measurements with GRAVITY data. Applying the analysis scheme developed here lifts any such discrepancies (see Sec.~\ref{sec: gc-previous}). In particular Fig.~\ref{fig: r0-fig} demonstrates that belatedly corrected data sets of earlier publications give fully consistent results whose accuracy increases with time.

\bibliographystyle{aa}
\bibliography{gravity-phasemaps}

\begin{appendix}

\section{List of Zernike coefficients}
\label{sec: zernike-coefficients}
The Zernike coefficients obtained by fitting the Calibration Unit measurements from late-2019 and early-2020 are summarized in Tabs. \ref{tab: zernike-ceofficients-2019} and \ref{tab: zernike-ceofficients-2020}, respectively. We provide the science channel results for all for GRAVITY beams (GV1 to GV4) in units of $\mu$m according to the definitions in Eqs. (\ref{eq: zernike-decomposition}) and (\ref{eq: zernike-fits-focal-plane-decomposition}), where $A_n^m$ labels pupil-plane aberrations and $B_n^m$ those in the focal plane.

\begin{table}
\begin{center}
\renewcommand{\arraystretch}{1.2}
\begin{tabular}{l|cccc}
 & GV1 & GV2 & GV3 & GV4 \\
 \hline
$A_0^0$& $-0.005$ & $-0.028$ & $-0.019$ & -0.014 \\
$A_1^{-1}$& $0.000$ & $0.008$ & $0.062$ & -0.014 \\
$A_1^1$& $0.021$ & $-0.030$ & $0.053$ & 0.022 \\
$A_2^{-2}$& $0.009$ & $-0.009$ & $0.028$ & 0.010 \\
$A_2^2$& $-0.010$ & $-0.012$ & $0.015$ & -0.035 \\
$A_2^0$& $-0.034$ & $-0.012$ & $-0.016$ & -0.002 \\
$A_3^{-1}$& $0.032$ & $-0.042$ & $0.028$ & 0.065 \\
$A_3^{1}$& $0.032$ & $0.071$ & $0.081$ & 0.013 \\
$A_3^{-3}$& $-0.056$ & $0.011$ & $0.032$ & 0.021 \\
$A_3^3$& $0.013$ & $-0.017$ & $-0.026$ & 0.054 \\
$A_4^{-2}$& $-0.005$ & $-0.020$ & $-0.036$ & -0.016 \\
$A_4^2$& $-0.049$ & $-0.014$ & $-0.046$ & -0.034 \\
$A_4^{-4}$& $0.011$ & $-0.005$ & $0.049$ & 0.002 \\
$A_4^4$& $-0.005$ & $-0.006$ & $-0.029$ & -0.012 \\
$A_4^0$& $-0.039$ & $-0.001$ & $-0.023$ & -0.019 \\
$A_5^{-1}$& $0.011$ & $0.030$ & $0.013$ & 0.014 \\
$A_5^1$& $-0.003$ & $-0.026$ & $0.032$ & -0.032 \\
$A_5^{-3}$& $0.018$ & $-0.026$ & $-0.015$ & -0.013 \\
$A_5^3$& $-0.020$ & $0.002$ & $-0.026$ & -0.030 \\
$A_5^{-5}$& $0.013$ & $-0.018$ & $0.008$ & -0.027 \\
$A_5^5$& $0.003$ & $-0.003$ & $0.047$ & -0.002 \\
$A_6^{-6}$& $-0.003$ & $0.009$ & $0.018$ & -0.001 \\
$A_6^6$& $-0.009$ & $0.013$ & $-0.019$ & 0.015 \\
$A_6^{-4}$& $-0.002$ & $0.004$ & $-0.017$ & 0.002 \\
$A_6^4$& $0.021$ & $0.000$ & $0.018$ & 0.018 \\
$A_6^{-2}$& $0.001$ & $-0.001$ & $0.002$ & -0.000 \\
$A_6^2$& $0.003$ & $0.002$ & $0.003$ & 0.002 \\
$A_6^0$& $0.024$ & $0.001$ & $0.024$ & 0.007 \\
$B_1^{-1}$& $0.010$ & $0.113$ & $0.065$ & 0.033 \\
$B_1^1$& $0.035$ & $-0.043$ & $0.062$ & 0.042 \\
$B_2^0$& $-0.006$ & $-0.011$ & $0.005$ & 0.007 \\
$B_2^{-2}$& $-0.045$ & $0.053$ & $-0.086$ & 0.024 \\
$B_2^2$& $0.011$ & $0.033$ & $-0.004$ & 0.031 \\
\end{tabular}
\end{center}
\caption{Zernike coefficients for science channel aberrations fitted to the calibration measurement on 03/11/19. All coefficient are given in units of $\mu\mathrm{m}$.}
\label{tab: zernike-ceofficients-2019}
\end{table}

\begin{table}
\begin{center}
\renewcommand{\arraystretch}{1.2}
\begin{tabular}{l|cccc}
 & GV1 & GV2 & GV3 & GV4 \\
 \hline
$A_0^0$& $-0.009$ & $-0.059$ & $-0.019$ & -0.027 \\
$A_1^{-1}$& $-0.018$ & $0.034$ & $0.066$ & -0.003 \\
$A_1^1$& $0.008$ & $0.016$ & $0.045$ & 0.043 \\
$A_2^{-2}$& $0.008$ & $-0.005$ & $0.047$ & 0.006 \\
$A_2^2$& $-0.012$ & $-0.010$ & $0.019$ & -0.023 \\
$A_2^0$& $-0.043$ & $-0.012$ & $-0.024$ & 0.012 \\
$A_3^{-1}$& $0.020$ & $-0.039$ & $0.038$ & 0.075 \\
$A_3^{1}$& $0.042$ & $0.079$ & $0.063$ & 0.026 \\
$A_3^{-3}$& $-0.031$ & $0.009$ & $0.029$ & 0.023 \\
$A_3^3$& $-0.001$ & $-0.006$ & $0.022$ & 0.032 \\
$A_4^{-2}$& $-0.028$ & $-0.049$ & $-0.042$ & -0.014 \\
$A_4^2$& $-0.030$ & $-0.052$ & $-0.019$ & -0.017 \\
$A_4^{-4}$& $0.014$ & $-0.014$ & $0.023$ & -0.014 \\
$A_4^4$& $-0.004$ & $-0.001$ & $-0.016$ & -0.016 \\
$A_4^0$& $-0.049$ & $-0.027$ & $0.001$ & -0.023 \\
$A_5^{-1}$& $0.022$ & $0.026$ & $-0.000$ & -0.000 \\
$A_5^1$& $-0.014$ & $-0.031$ & $0.034$ & -0.041 \\
$A_5^{-3}$& $0.005$ & $-0.027$ & $-0.011$ & -0.017 \\
$A_5^3$& $-0.007$ & $-0.005$ & $-0.025$ & -0.017 \\
$A_5^{-5}$& $0.004$ & $-0.015$ & $0.008$ & -0.007 \\
$A_5^5$& $-0.008$ & $0.001$ & $0.058$ & 0.004 \\
$A_6^{-6}$& $-0.006$ & $0.018$ & $0.040$ & 0.014 \\
$A_6^6$& $0.001$ & $0.008$ & $-0.002$ & 0.008 \\
$A_6^{-4}$& $0.013$ & $0.017$ & $-0.005$ & 0.001 \\
$A_6^4$& $0.012$ & $0.021$ & $0.014$ & 0.015 \\
$A_6^{-2}$& $-0.001$ & $0.002$ & $0.003$ & 0.001 \\
$A_6^2$& $-0.001$ & $-0.001$ & $0.006$ & 0.004 \\
$A_6^0$& $0.030$ & $0.007$ & $0.016$ & 0.009 \\
$B_1^{-1}$& $0.002$ & $0.115$ & $0.036$ & 0.023 \\
$B_1^1$& $0.068$ & $-0.032$ & $0.086$ & 0.035 \\
$B_2^0$& $-0.004$ & $-0.000$ & $0.004$ & 0.015 \\
$B_2^{-2}$& $-0.027$ & $0.035$ & $-0.076$ & 0.012 \\
$B_2^2$& $0.043$ & $0.065$ & $-0.040$ & 0.008 \\
\end{tabular}
\end{center}
\caption{Zernike coefficients for science channel aberrations fitted to the calibration measurement on 03/03/20. All coefficient are given in units of $\mu\mathrm{m}$.}
\label{tab: zernike-ceofficients-2020}
\end{table}

\section{Data}
\label{sec: gc-data}
We use the data set presented in \cite{2020A&A...636L...5G} with the following changes:
\begin{itemize}
\item Each single-beam astrometric position is corrected according to Eq.~(\ref{eq: s2-correction}), and we add the statistical error of this correction in quadrature, which increases the individual uncertainties by around $15\,\mu$as.
\item We corrected the radial velocity of the epoch 2018.1277, which was $13\,$km/s too high in the previous data set.
\item Further, we are able to add one interferometric position measurement of S2 from early March 2020. Like in 2019, the separation between S2 and Sgr~A* exceeds the fiber field of view, and hence a dual-beam measurement needed to be employed.
\end{itemize}
Our data set consists of 128 AO-based astrometric points, 58 GRAVITY-based astrometric points and 97 radial velocities, of which the first three before 2003 are from \citet{2019Sci...365..664D}.

\subsection{Dual-beam measurement in 2020}

Due to the limited observability of the GC in early March and expecting observations in the following months, we did not attempt to observe Sgr~A* in March 2020, but only pointed to S2 and to our usual calibrator star R2, with the aim of testing the stability of the GRAVITY astrometry. Pointings to Sgr~A* were planned for later in the year. They had to be canceled due to the pandemic-related closure of the VLT(I) from mid-March on. To still determine the S2 -- Sgr~A* separation vector from this observation, we need to proceed in two steps and first measure the S2 -- R2 distance, then we reference R2 to Sgr~A*.

The distance between S2 and R2 is measured with the dual-beam method (Sec.~\ref{sec: s2-corrections}), where we calibrate the S2 files with R2. In addition to the 2020 measurement, this separation is also available for 56 epochs in the years 2017, 2018 and 2019. It can be measured very precisely due to the brightness of the two stars. Since the S2 -- Sgr~A* vectors have already been determined in \cite{2020A&A...636L...5G}, we can also refer R2 to Sgr~A* in those earlier epochs. We then fit a simple quadratic function for the time evolution of the R2 coordinates relative to Sgr~A* and extrapolate it to March 2020. Given the large number of data and the small time range to extrapolate for, the extra uncertainty introduced is well below the $100\,\mu$as level.

We derive the S2 position in 2020 from the four scientifically usable exposures as their mean. We assign an error of $150\,\mu$as to each coordinate for this data point, reflecting both the smaller number of files compared to what we typically had available in 2019 and the extra uncertainty due to the additional step of referencing via R2. The new data point falls well onto the expected orbit, but its error bar is too large to have a significant impact on the fitted parameters. 

\section{Analysis of the difference between $R_0$ determinations from Keck and VLT data sets}
\label{sec:AppA}
While we believe our determination of $R_0$ is accurate to the $40\,$pc level, we note that the value published in \citet{2019Sci...365..664D} is discrepant at the $300\,$pc level. Both teams use the orbit of the star S2 around Sgr~A* for the $R_0$ determination, and hence it is natural to ask where the differences are.
\subsection{Data}
\label{sec:AppA1}
Beyond our ("VLT") data set (App.~\ref{sec: gc-data}), we use the Keck data set published in \citet{2019Sci...365..664D}. We apply the NIRC2 radial velocity offset of $+80\,$km/s as determined in \citet{2019Sci...365..664D} to the NIRC2 data, i.e. we add 80 km/s to these radial velocities. Unlike \citet{2019Sci...365..664D}, we then don't fit for this offset.
Further, we drop the last astrometric data point (epoch 2018.67148268), as suggested by the authors in a private communication. The data set consists of 45 astrometric points and 116 radial velocities, of which 41 are actually from the VLT data set between 2003 and 2016. The published table also includes one radial velocity from the epoch 2019.3567, which possibly was not part of the data set actually used in \citet{2019Sci...365..664D}.

\subsection{The difference in $R_0$}
\label{sec:AppA2}
We fit the orbit with a simple, 13-parameter model: The six orbital elements of the star (corresponding to the initial conditions of the star in phase space), six parameters for the position and velocity of the MBH, and the mass of the MBH. The fits are done using the relativistic corrections as in \citet{2020A&A...636L...5G}, i.e. we fix $f_\mathrm{RS} = f_\mathrm{SP} = 1$. For this non-Keplerian motion, the meaning of the orbital elements is that they are osculating at a reference epoch, for which we choose T=2010.35, close to the apocenter passage time of S2.

For fitting the VLT data set, we use the same approach as in \citet{2020A&A...636L...5G}: For the GRAVITY data, we assume that the astrometry directly refers the S2 positions to the mass center, as we directly measure the separation vector between the two objects interferometrically. For the NACO (AO-imaging based) data, we allow for a coordinate system offset, on which we set priors following the work from \citet{2015MNRAS.453.3234P}, and we include the NACO flare positions as an additional constraint for locating the mass. This fit yields
\begin{eqnarray} \nonumber
R_0 &=& 8274.9 \pm 9.3\, \mathrm{pc}\\ \nonumber
a &=& 124.982 \pm 0.034\, \mathrm{mas}\\ \nonumber
i &=& 134.685 \pm 0.029^\circ\\ 
\Omega &=& 227.175 \pm 0.029^\circ  ,
\label{res_mpe}
\end{eqnarray}
where $a$ is the semi-major axis, $i$ the inclination and $\Omega$ the position angle of ascending node of the S2 orbit, and the errors are the statistical fit uncertainties. The VLT astrometry is dominated by the GRAVITY points, as illustrated by dropping all AO data points, which results in $R_0 = 8276 \pm 10\,$pc. 

Fitting the Keck data set with the same 13-parameter model as used for Eq.~\ref{res_mpe} yields 
\begin{eqnarray} \nonumber
R_0 &=& 7935 \pm 44\, \mathrm{pc}\\ \nonumber
a &=& 126.64 \pm 0.27\, \mathrm{mas}\\ \nonumber
 i &=& 133.78 \pm 0.15^\circ\\ 
 \Omega &=& 227.66 \pm 0.13^\circ  .
\label{res_ucla}
\end{eqnarray}
This is not the exact same number as in \citet{2019Sci...365..664D}, where $R_0 = 7959 \pm 59\,$pc is reported. The small (and statistically insignificant) difference is most likely due to the noise model which \citet{2019Sci...365..664D} include in their analysis, which we do not have readily available. Applying the noise model at hand \citep{2018MNRAS.476.4372P, 2019A&A...625L..10G} yields $R_0 = 7965 \pm 56\,$pc. Hence, the value reported by \citet{2019Sci...365..664D} lies between the two numbers we get by re-fitting their data. In the following, we will use for simplicity, and for equal treatment of the data, the value and approach as in Eq.~\ref{res_ucla}. We have thus a difference of $\Delta R_0 = 340 \pm 45\,$pc.

\subsection{Comparing, combining \& adjusting the astrometry}
\label{sec:AppA3}
Already, \citet{2009ApJ...707L.114G} noticed that a simple attempt to compare the astrometric data sets by plotting them on top of each other fails. One needs to allow for an offset and a drift between 
the two coordinate systems (i.e. four parameters $\Delta x,\,\Delta y,\, \Delta v_x,\,\Delta v_y$). This yields thus a 17-parameter fit. Comparing the best-fitting parameters in Eq.~\ref{res_mpe} and Eq.~\ref{res_ucla} shows that they differ in $\Omega$ significantly. This parameter is fully degenerate with the angular orientation (called $\beta$ here) of the coordinate system. Hence, the difference in $\Omega$ suggests that the two astrometric data sets are rotated with respect to each other.

Therefore we extend the combination scheme by an additional, fifth parameter, $\Delta \beta$, resulting in a 18-parameter fit. With this we fitted both data sets simultaneously, omitting the 41 VLT radial velocities from the Keck data set, whist dropping also the three Keck ones in the VLT data set. This fit matches the two coordinate systems ideally onto each other and results in
\begin{eqnarray} \nonumber
R_0 &=& 8260 \pm 9\, \mathrm{pc}\\ \nonumber
a &=& 125.00 \pm 0.03\, \mathrm{mas}\\ \nonumber
i &=& 134.66 \pm 0.03^\circ\\ \nonumber
\Omega &=& 228.16 \pm 0.03^\circ\\
\Delta \beta &=&0.32 \pm 0.05^\circ  ,
\label{res_comb}
\end{eqnarray}
Note that the value of $\Delta \beta$ matches the difference $\Delta \Omega$. We conclude that indeed the Keck and VLT data are rotated with respect to each other. The other parameters are 
very similar to Eq.~\ref{res_mpe}, which is due to the considerably smaller astrometric uncertainties of the GRAVITY data compared to the adaptive optics data.

With the best-fit coordinate system difference in hand, we can transform the Keck astrometric data into the VLT coordinate system and vice versa. We choose to do the former, since the VLT data set is more directly calibrated by the interferometric data. After applying the coordinate system difference to the Keck data, we can fit them again with a 13-parameter model. This yields the exact same best-fit parameters as in Eq.~\ref{res_ucla} (with the exception of $\Omega$, of course). Hence, transforming the astrometry does not change the more fundamental differences between the two orbits, while a direct comparison is now feasible. The value of $\Omega$ can be omitted in the following.

\subsection{Discrepancy in the radial velocity data}
\label{sec:AppA4}
   \begin{figure*}
   \centering
  \includegraphics[width=0.41\linewidth]{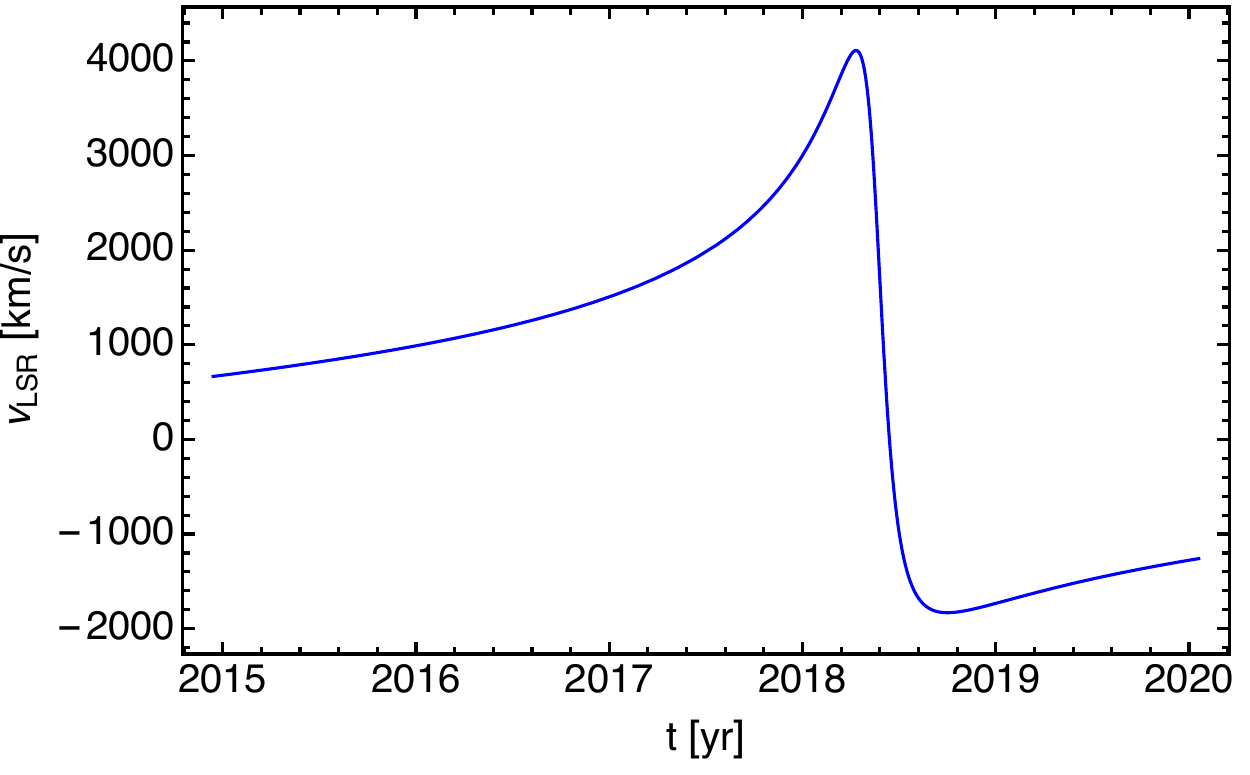}
  \includegraphics[width=0.4\linewidth]{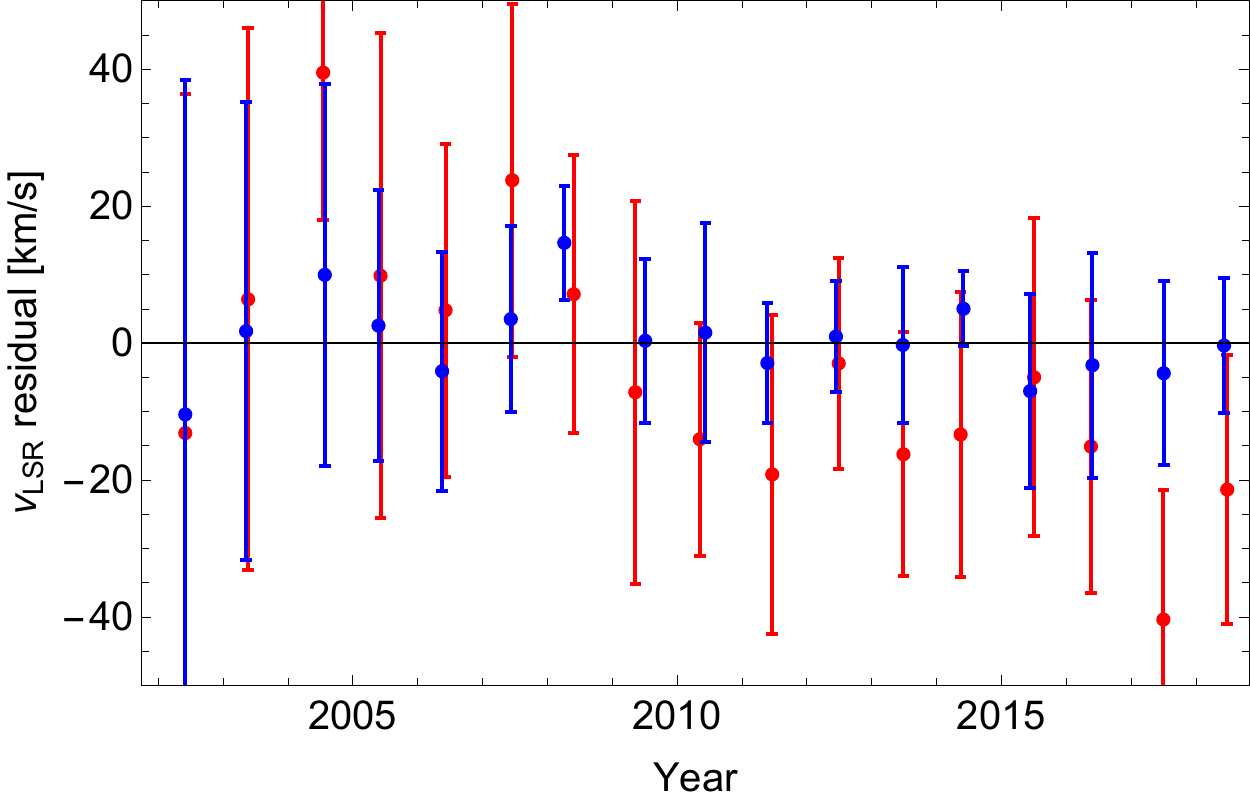}
  \includegraphics[width=0.4\linewidth]{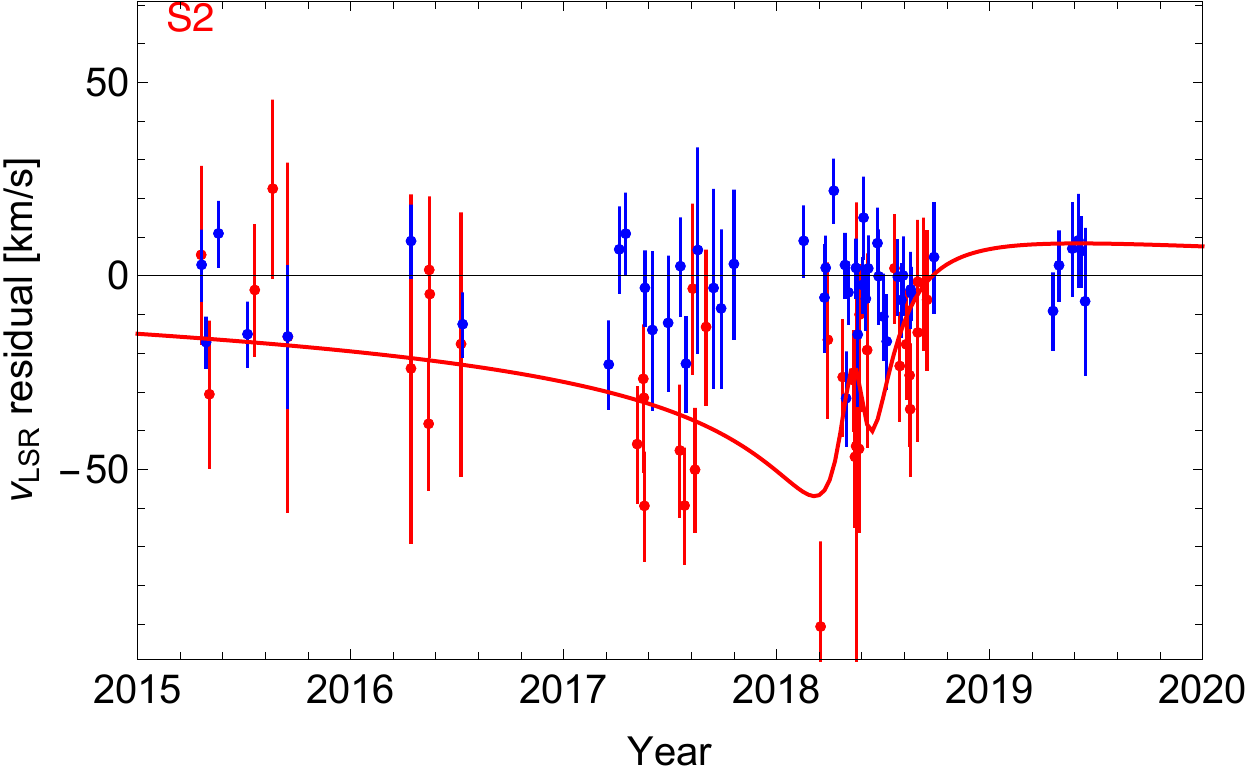}
   \includegraphics[width=0.4\linewidth]{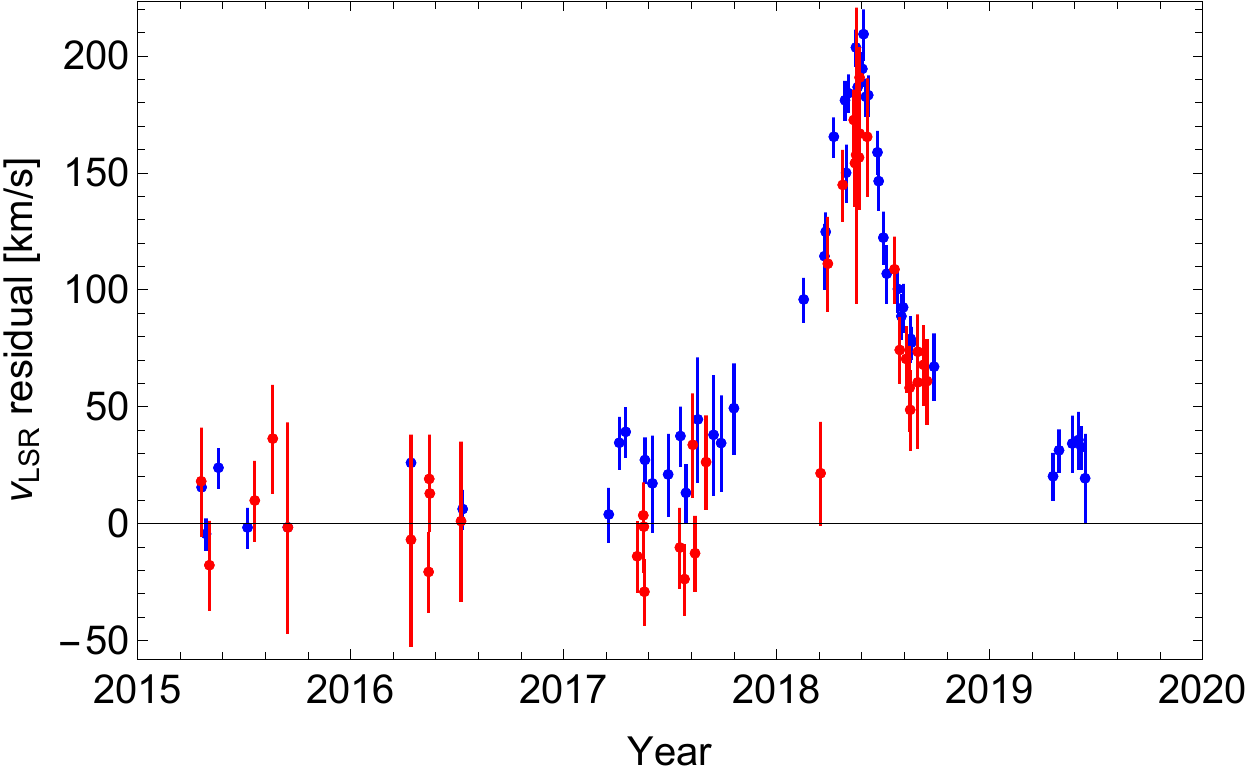}
     \caption{\small Comparison of the radial velocity data sets. Blue points are data from the VLT data set, red from the Keck data set. Top left: Radial velocity as a function of time for the VLT fit (Eq.~\ref{res_mpe}). Top right: Yearly averages of the residua of the two data sets to the fit from Eq.~\ref{res_mpe}. By construction the VLT data thus scatter around 0. The Keck data deviate systematically from 2011 on, and the discrepancy increases in the later years. Bottom left: The same as the left panel, but zooming in to the period 2015 - 2020, and showing all individual data points. The best fit Keck orbit corresponding to Eq.~\ref{res_ucla} is the red line. Apparently, the difference is largest, when the radial velocity gets largest (in the year 2018 at pericenter passage). Bottom Right: Both data sets show a clear peak in radial velocity in 2018 when comparing with the Keplerian part of the VLT fit (Eq.~\ref{res_mpe}), i.e. both data sets clearly detect the redshift term. }
     \label{app1fig1}
    \end{figure*}

\citet{2018ApJ...854...12C} have investigated the consistency of the radial velocity data between the Keck and VLT data sets for the years 2000 to 2016, and they concluded that the data are in agreement with each other. We have repeated the exercise, now also extending into the time of the pericenter passage in 2018 (Fig.~\ref{app1fig1}). To our surprise, the radial velocities 
differ systematically from $\approx 2011$ on, and the difference gets larger as the radial velocity increases ever more. The difference reaches $\approx 50\,$km/s in 2018, just before the star swung through pericenter \footnote{Also, there is one obvious outlier in the Keck data, the earliest 2018 point. We have checked that dropping this measurement does not change the Keck-fit result in any significant way.}. 

Hence, it is an obvious question to ask what influence the radial velocities have on $R_0$? For this, we swapped the radial velocities between the two data sets. Using the VLT-set together with the Keck astrometry yields
\begin{eqnarray} \nonumber
R_0 &=& 8094 \pm 32\, \mathrm{pc}\\ \nonumber
a &=& 126.08 \pm 0.21\, \mathrm{mas}\\ 
i &=& 134.0 \pm 0.13^\circ
\label{res_1}
\end{eqnarray}
Vice versa, using the Keck radial velocities together with the VLT astrometry yields $R_0 = 8214 \pm 14\, \mathrm{pc}$. Given that the Keck radial velocity set contains 35\% VLT radial velocities, the fit in Eq.~\ref{res_1} is the cleaner test. We thus explain roughly half of the difference in $R_0$ with the radial velocity data, i.e. $159\,$pc.

Why do the radial velocities differ? So far, we can only offer an explanation for $\approx$ 20\% of the radial velocity difference: We applied the stellar atmosphere model-based fitting with the StarKit package used in \citet{2019Sci...365..664D} also to the VLT spectroscopy. We found a significant difference for large radial velocities, which we were able to trace down to the Doppler formula used by the StarKit package. While both \citet{2019Sci...365..664D} and \citet{2020A&A...636L...5G} state that the spectroscopic observable is $v_r = z\,c$, i.e. the redshift of a given spectrum, the StarKit package actually applies a Doppler formula which includes the longitudinal, relativistic correction: $\lambda' = \lambda_0 \sqrt{\frac{1+v_r/c}{1-v_r/c}}$. In this form, the Doppler formula ignores the (significant) tangential motion $v_t$ of S2. In order to apply a relativistic correction one needs to use the full Doppler formula $1+z = \frac{1+v_r/c}{\sqrt{1-(v_r^2+v_t^2)/c^2}}$ \citep{2003A&A...401.1185L}. For this correction, however, the spectroscopic information is not sufficient. One cannot, in general, Doppler-correct a spectrum in a relativistic way without knowing the other motion component. Further, even if one would apply the full correction, one would in the following of course not be able to fit for the relativistic redshift anymore. 

The difference between the two formulae is small at velocities much smaller than the speed of light, but becomes important close to peri-center, when S2 reaches a velocity of nearly $8000\,\mathrm{km/s}$. Still, it amounts to $\approx 25\,\mathrm{km/s}$ at most and thus is smaller than the observed difference in Fig.~\ref{app1fig1}. This difference is also visible in Fig.~1 of \citet{2019Sci...365..664D}: The plotted model spectra are slightly more redshifted than what the underlying data suggest. Changing the Keck radial velocities accordingly yields a fit with $R_0 = 7972 \pm 44\,$pc, i.e. accounting for $37\,$pc of the $159\,$pc.

Further checks did not yield any clues why there remains a significant difference in the radial velocities. We note:
\begin{itemize}
\item We checked whether the time stamps are assigned consistently between the two data sets, and did not find a difference.
\item Fig.~\ref{app1fig1} bottom right shows that both data sets clearly show the redshift peak around pericenter.
\end{itemize}

\subsection{Discrepancy in the astrometry}
\label{sec:AppA5}
   \begin{figure*}
   \centering
  \includegraphics[width=0.41\linewidth]{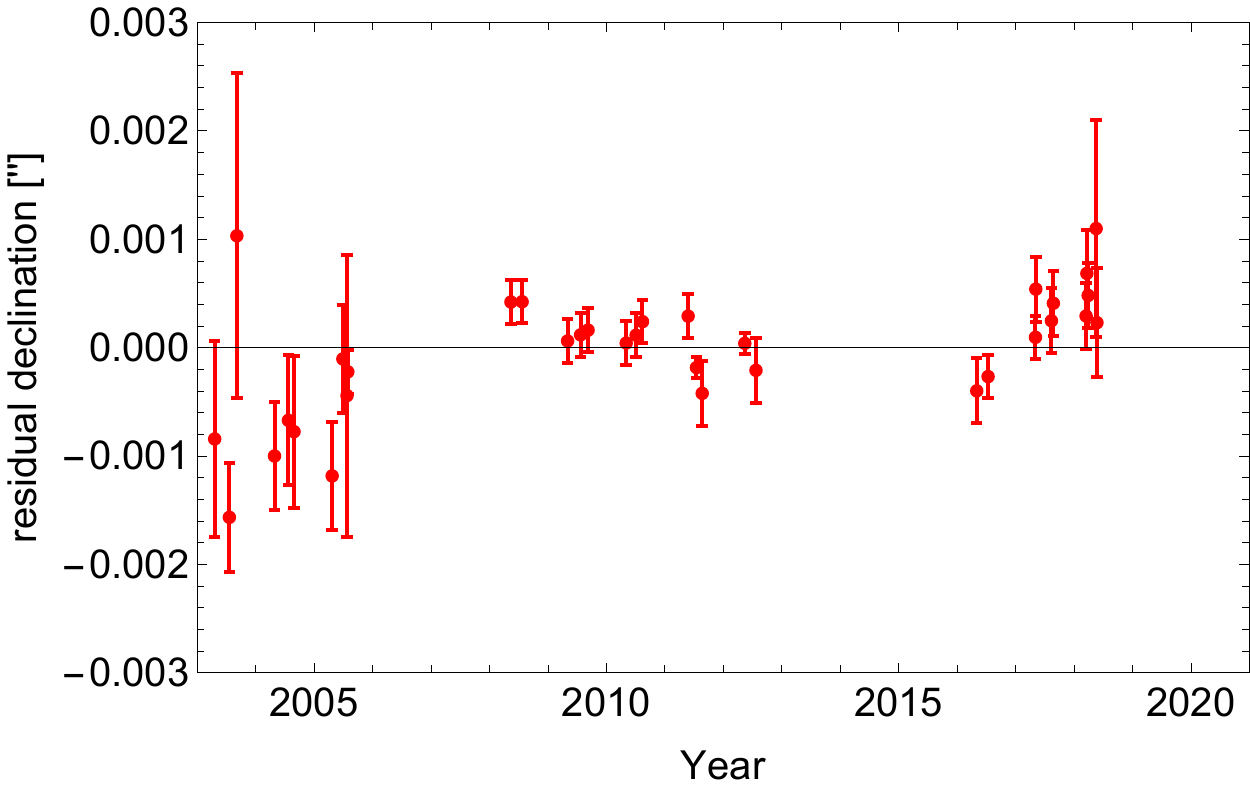}
  \includegraphics[width=0.4\linewidth]{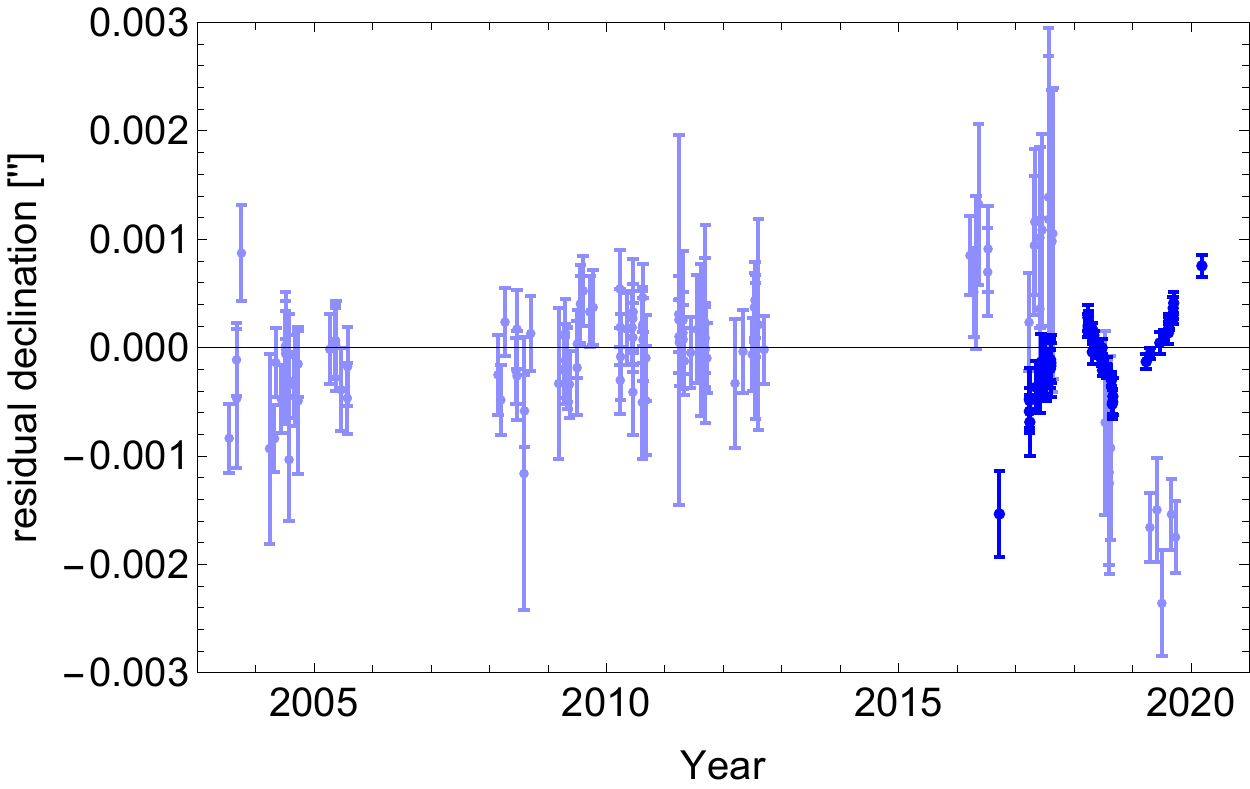}
     \caption{\small Comparison of the astrometric residual after forcing an offset in declination such that the fit to Keck data set matches the VLT one (left) and such that the fit to VLT data set matches the Keck one (right). Lighter blue corresponds to AO data from the VLT data set, darker blue to the GRAVITY data.}
     \label{app1fig2}
    \end{figure*}

Comparing the fits in Eq.~\ref{res_mpe} and Eq.~\ref{res_ucla} shows that they not only differ in $R_0$, but also in the size of the semi-major axis $a$. We find $\Delta a/a = 1.28 \pm 0.22\,\%$. The same is not true for the semi-minor axis though, $\Delta b/ b$ is consistent with 0. Interestingly, the projected ellipses as given by the astrometric data in the plane of sky agree in both semi-major and semi-minor axes to within $0.17\%$. Hence, the inclinations $i$ need to differ, which Eq.~\ref{res_mpe} and Eq.~\ref{res_ucla} confirm. We find in accordance with the above $1 \, - \, \sin(i_\mathrm{VLT}) / \sin(i_\mathrm{Keck}) \approx 1.3\%$. 

The inclination of the ellipse determines where the projected center of mass is located. Given the orientation of the S2 orbit and the disagreement in $a$ but not in $b$ hints  towards an offset of the center of mass in the declination direction. Indeed, we can show that introducing an offset to either $y$ or $v_y$ (the mass position and velocity in declination) can explain the remaining discrepancy. Starting from the fit of the transformed Keck data set, we fix $v_y$ to its best fit value of $-0.15\,$mas/yr. All other parameters are left free again for a subsequent fit. Additionally using the VLTI velocities in this fit instead of the Keck ones yields:
\begin{eqnarray} \nonumber
R_0 &=& 8277 \pm 28\, \mathrm{pc}\\ \nonumber
a &=& 124.76 \pm 0.16\, \mathrm{mas}\\ 
i &=& 134.63 \pm 0.11^\circ\,.
\label{res_4}
\end{eqnarray}
This fit yields thus from the Keck astrometry the same value for $R_0$ as the VLT fit. Also note, that indeed semi-major axis $a$ and inclination $i$ have moved to the VLT values by forcing $v_y$ to have an offset. Since the mass position is parametrized with a time origin at T=2000.0, the best fit $y$ also changes, from $-0.972\,$mas to $1.234\,$mas. The systematic uncertainty on $y$ and $v_y$ estimated by \citet{2019Sci...365..664D} are $1.16\,$mas and $0.066\,$mas/yr respectively. Hence, the difference one needs to enforce is within $\approx2\sigma$ of the systematic uncertainty, and the residuals in fig.~\ref{app1fig2} (left) appear to be acceptable. Essentially the same can be achieved by forcing an offset to $y$ and leaving $v_y$ free instead. 

Can one can turn the argument around and apply a similar offset to the VLT data in order to lower the VLT-based value of $R_0$? In a first attempt we applied the same offset to the VLT AO data. However, even an offset 10 x larger (i.e. $1.2\,$mas/yr), changes $R_0$ only by $\approx30\,$pc. This is not surprising, since the VLT astrometry is completely dominated by the GRAVITY data.
Thus, we instead tried varying $v_y$ and $y$ for the GRAVITY data, giving up the assumption that the GRAVITY source directly is the mass center. Also, we exchanged the VLT radial velocities for the Keck ones. We find that we need to change $v_y$ by  $-1.4\,$mas/yr in order to get a distance similar to the Keck value:
\begin{eqnarray} \nonumber
R_0 &=& 7928 \pm 16\, \mathrm{pc}\\ \nonumber
a &=& 126.89 \pm 0.05\, \mathrm{mas}\\ 
i &=& 133.51 \pm 0.03^\circ\,.
\label{res_6}
\end{eqnarray}
The fit achieves the lower $R_0$ by tilting the orbit similar to the fit from Eq.~\ref{res_ucla}. The enforced change of $v_y$ is unrealistically large ($12\times$ larger than what was needed for the Keck data), Also, the GRAVITY data show very strong and systematic residuals of up to $0.5\,$mas (fig.~\ref{app1fig2} right), and the reduced $\chi^2$ of the fit increased from $1.50$ to $2.63$.

\end{appendix}

\end{document}